\documentclass[12pt]{article}

\newenvironment{wileykeywords}{\textsf{Keywords:}\hspace{\stretch{1}}}{\hspace{\stretch{1}}\rule{1ex}{1ex}}

\usepackage{amsmath,amssymb}
\usepackage{graphicx}
\usepackage{color}
\usepackage{dcolumn}
\usepackage{bm}
\usepackage[numbers,super,comma,sort&compress]{natbib}
\usepackage{booktabs, array, multirow}
\usepackage{subcaption}

\definecolor{background-color}{gray}{0.98}

\title{The Adaptive Buffered Force QM/MM method in the CP2K and AMBER software
packages}

\author{Letif Mones \thanks{Engineering Department, University of Cambridge, Cambridge, CB2 1PZ, United Kingdom}, Andrew Jones \thanks{School of Physics and Astronomy, University of Edinburgh EH9 3JZ, United Kingdom}, Andreas W. G\"otz \thanks{San Diego Supercomputer Center, University of California San Diego, La Jolla, California 92093, United States}, Teodoro Laino \thanks{IBM Research--Zurich, S\"aumerstrasse 4, 8803 R\"uschlikon, Switzerland}, \\
Ross C. Walker \footnotemark[3] \thanks{Department of Chemistry and Biochemistry, University of California San Diego, La Jolla, California 92093, United States}, Ben Leimkuhler \thanks{The Maxwell Institute and School of Mathematics, University of Edinburgh EH9 3JZ, United Kingdom}, G\'abor Cs\'anyi \footnotemark[1], Noam Bernstein \thanks{Naval Research Laboratory, Center for Computational Materials Science, Washington, DC 20375, United States of America}}

\begin{document}

\maketitle

\begin{abstract}
The implementation and validation of the adaptive buffered force QM/MM method in
two popular packages, CP2K and AMBER are presented. The implementations build on
the existing QM/MM functionality in each code, extending it to allow
for redefinition of the QM and MM regions during the simulation and
reducing QM-MM interface errors by discarding forces near the
boundary according to the buffered force-mixing approach.  New
adaptive thermostats, needed by force-mixing methods, are also
implemented. Different variants of the method are benchmarked by
simulating the structure of bulk water, water autoprotolysis
in the presence of zinc
and dimethyl-phosphate hydrolysis using various semiempirical Hamiltonians
and density functional
theory as the QM model. It is shown that with suitable parameters, based on force
convergence tests, the adaptive buffered-force QM/MM scheme can
provide an accurate approximation of the structure in the dynamical
QM region matching the corresponding fully QM simulations, as well
as reproducing the correct energetics in all cases.  Adaptive
unbuffered\ force-mixing and adaptive conventional QM/MM methods
also provide reasonable results for some systems, but are more
likely to suffer from instabilities and inaccuracies.  \end{abstract}

\begin{wileykeywords}
QM/MM, adaptive QM/MM, force-mixing, multi scale
\end{wileykeywords}

\clearpage

\begin{figure}[h]
\centering
\colorbox{background-color}{
\fbox{
\begin{minipage}{1.0\textwidth}
\includegraphics[width=50mm]{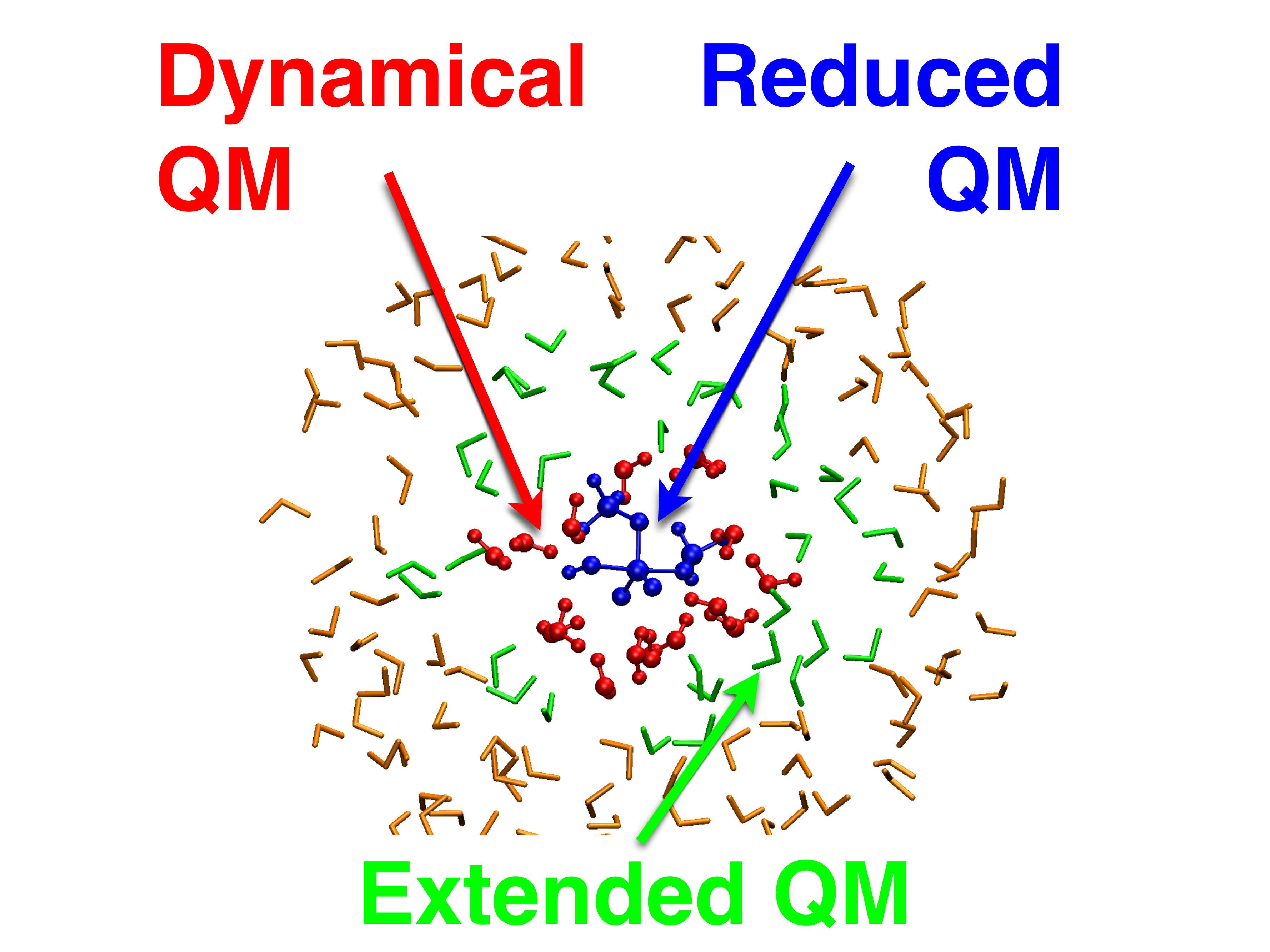} 
\\
We present implementations of an adaptive method for QM/MM simulations
in the AMBER and CP2K  packages which make it straightforward
to  describe quantum mechanically not only the reacting species, but also a surrounding region
of solvent, because the set of quantum atoms can be changed at will during the simulation.
We compare geometries and free energy profiles to
those of fully quantum mechanical simulations and show that our scheme
is more robust than alternatives.

\end{minipage}
}}
\end{figure}

\makeatletter
\renewcommand\@biblabel[1]{#1.}
\makeatother

\bibliographystyle{apsrev}

\renewcommand{\baselinestretch}{1.5}
\normalsize

\clearpage

\section*{\sffamily \Large INTRODUCTION}

Quantum-mechanics/molecular-mechanics (QM/MM) methods\cite{Warshel76}
have matured over the past few decades and are now an essential tool for modeling chemical
reactions of complex systems.  Most of the system is described by
typically non-reactive MM force fields, but these give a
poor (or even no) description of chemical processes such as changing
charge state or covalent bond rearrangement.   A quantum mechanical description is
used in the  region
where such processes occur, and the extent of this region is kept small due to the associated computational expense.
The QM and MM subsystems affect each other directly, by covalent,
electrostatic, or other non-bonded interactions, as well as implicitly
through long-range structure in the MM subsystem.  Capturing such long range interactions can be essential even for the
description of the local structure, e.g.\ a in protein where the reaction
  involves residues that are kept in place by the
structure of the rest of the protein, or because
long range electrostatic effects play a direct role in the reaction.\cite{Villa01, Senn09}

For a QM/MM method to describe the complete system accurately, the
individual methods used for the QM and MM descriptions must be
appropriate for the configurations and processes in their respective
regions, and the interaction between them must be accounted for.  The dominant approach, which  we will call \textit{conventional}
QM/MM (conv-QM/MM), is to fix the set of atoms in the QM and MM
subsystems and define the total energy of the system as a sum of the QM energy of the QM region, the MM energy of the MM
region, and an interaction energy.  The interaction term can include
the non-bonded and electrostatic energies of MM descriptions of the
QM atoms in the field of the MM atoms (``mechanical
embedding''),\cite{Bakowies96} or it may include the
effect of the MM electrostatic field on the  QM description,
including  the explicitly described electron density
(``electrostatic embedding'').\cite{Bakowies96} If
covalent bonds across the QM-MM interface are present, they must
be capped in some way in the QM description so as to eliminate dangling bonds in the QM
subsystem, e.g.\ using H atoms\cite{Singh86}, generalized hybrid
orbitals\cite{Gao98} or pseudopotentials\cite{Zhang99}. The accuracy of the conventional
approach depends on the appropriateness of using a fixed set of
atoms in the QM region, and on the ability of the QM-MM interaction
term to eliminate the fictitious boundary effects in the QM and MM subsystem calculations.

Carrying out QM/MM simulations on different sized QM regions shows
that widely used interaction terms lead to significant
errors in the atomic forces near the QM-MM
interface when compared to calculations using very large QM regions or describing the entire system quantum mechanically  using periodic boundary conditions
(we will refer to the latter as ``fully QM'').\cite{AkinOjo:2008eh,Solt09,Hu11,AkinOjo:2011ip}
Although in many cases the effect on relevant observables can be small, these errors can be very
 problematic when the set of QM  atoms
is allowed to change.  In such adaptive methods,\cite{Kerdcharoen96, Kerdcharoen02, Rode05, Heyden07, Heyden08, Bulo09, Yockel11, Bulo13} which are
used to enable the QM region to move or  species to diffuse  in or
out of the reaction site, errors near the interface can lead to an
instability and a  net flux of atoms between the QM and MM regions
resulting in  unphysical density variations.\cite{Bernstein12, Varnai13}

There are a number of fundamental issues that must be addressed in
the design of any method that couples different descriptions in different regions
of a single system.  The way they are addressed can have particular
implications for adaptive  simulations, which may be different from
the way the choices affect simulations where the set of atoms in each subsystem is fixed.  One choice is whether the coupling
is formulated in terms of energy\cite{Kerdcharoen02, Heyden07, Ensing07, Heyden08, Yockel11, Bulo13} 
or forces\cite{Kerdcharoen96, Csanyi04, Rode05, Praprotnik05, Praprotnik06, Bulo09, Bernstein12, Varnai13}.  If it is formulated
in terms of energy, the total energy of the coupled system can be
defined, and changes of that energy as atoms or molecules switch between
descriptions can adversely affect the simulation.  This  can be
represented as a difference in chemical potential of the switching species 
being described with the two models.  A mismatch at any point
in space  for any molecular conformation will lead to unphysical
forces on atoms as they switch description, leading to transport of atoms
to the lower chemical potential region.  Coupling in terms of forces
can avoid this chemical potential mismatch effect, at the cost of
forgoing energy conservation  because no total energy can be
defined, due to the non-conservative nature of the  forces used to drive the dynamics.  
This tradeoff motivated the choice to use a force-based approach
both in our work, the Hot Spot~\cite{Kerdcharoen96} 
and difference-based adaptive solvation (DAS)~\cite{Bulo09} methods.   The use of non-conservative forces would
lead to unstable molecular dynamics trajectories, which we avoid by  using adaptive 
thermostats. These  have been shown to sample the correct distribution even in the presence of net heat generation.~\cite{Jones11}  

Another choice is whether the transition between the two descriptions
is abrupt or continuous.  An abrupt transition leads to discontinuities
in the dynamics as atoms suddenly switch from one region to another.  Employing a  transition region can make the energy or forces continuous by
smoothly interpolating between multiple calculations, but increases the
number of force calculations that must be performed.  While many published methods use transition regions
to smooth out such switching discontinuities,\cite{Heyden07, Heyden08, Kerdcharoen02, Yockel11, Kerdcharoen96, Rode05, Bulo09, Bulo13, Park12}
we have found that using abrupt transitions within a force-mixing approach
does not seem to significantly
affect the accuracy of average structures and free energy profiles.~\cite{RepProgPhys, Bernstein12, Varnai13}

The third choice is how the errors near the interface between the
two regions are handled.  Energy based methods are formulated in
terms of an MM energy, a QM energy, and the interaction term, and the
accuracy of the last one determines this error.  Adaptive methods
like ONIOM-XS~\cite{Kerdcharoen02} and SAP~\cite{Heyden07, Heyden08}
simply combine a weighted sum of several such calculations, and
therefore include a weighted sum of interface related errors.  Methods that mix a quantity
that can be localized to each atom can, in general, improve on this
using buffers, as we explain below.   Because the energy, especially
in the QM description, can not be localized to each atom, such
mixing is generally applied to forces.~\cite{Kerdcharoen96, Rode05, Bulo09, Bulo13}
The buffer regions used to improve boundary
force errors
are conceptually distinct from the transition regions mentioned above that
help  smooth discontinuities.

Over the past few years we have developed the adaptive buffered-force QM/MM method (adbf-QM/MM),
which uses force-mixing, abrupt transitions, and buffers to reduce
the effect of interface errors and enable stable adaptive
simulations.~\cite{Bernstein12} Many other published methods can also
be characterized in terms of the above choices.
The ABRUPT method\cite{Bulo13} is equivalent to a conv-QM/MM simulation,
using energy based coupling, where atoms switch abruptly between
the two descriptions without buffers.  The Hot Spot
method~\cite{Kerdcharoen96} uses force-mixing with transitions that
are interpolated over a region of about 0.5~{\AA}, but no buffers.
Sorted adaptive partitioning (SAP)~\cite{Heyden07},
ONIOM-XS~\cite{Kerdcharoen02}, and difference-based adaptive solvation
(DAS)~\cite{Bulo09} all use smooth transitions and no buffers, but
the first two use an energy based coupling while the last uses
force-mixing.  The SAP and DAS methods  require one calculation per
molecule in the  transition region, and the ONIOM-XS
method is limited to a single molecule in that region.

In previous publications we tested the adbf-QM/MM method on the
structure of bulk water,\cite{Bernstein12} as well as the free
energy profiles of two reactions in water, nucleophilic substitution
in methyl chloride and the deprotonation of tyrosine.\cite{Varnai13}
Here we describe the implementation of the adbf-QM/MM method in two
popular software packages, CP2K\cite{cp2k} and AMBER.\cite{amber, Salomon13}
The implementation extends the  QM/MM capabilities of the packages,
and with appropriate choice of parameters can be used to carry out  adaptive QM/MM simulation with or without buffering and force-mixing.
We test the different variants  using a variety of
QM models, including density functional theory (DFT) and semi-empirical
quantum mechanical (SE) models, on the structure of bulk water,  the free energy
profiles of dimethyl-phosphate hydrolysis and the autoprotolysis of water in the presence of a zinc ion.

\section*{\sffamily \Large METHODOLOGY}
\label{sec:methods}

\subsection*{\sffamily \large Overview of Adaptive Buffered Force QM/MM method}

\begin{figure}
        \centering
        \includegraphics[width=0.8\columnwidth, clip]{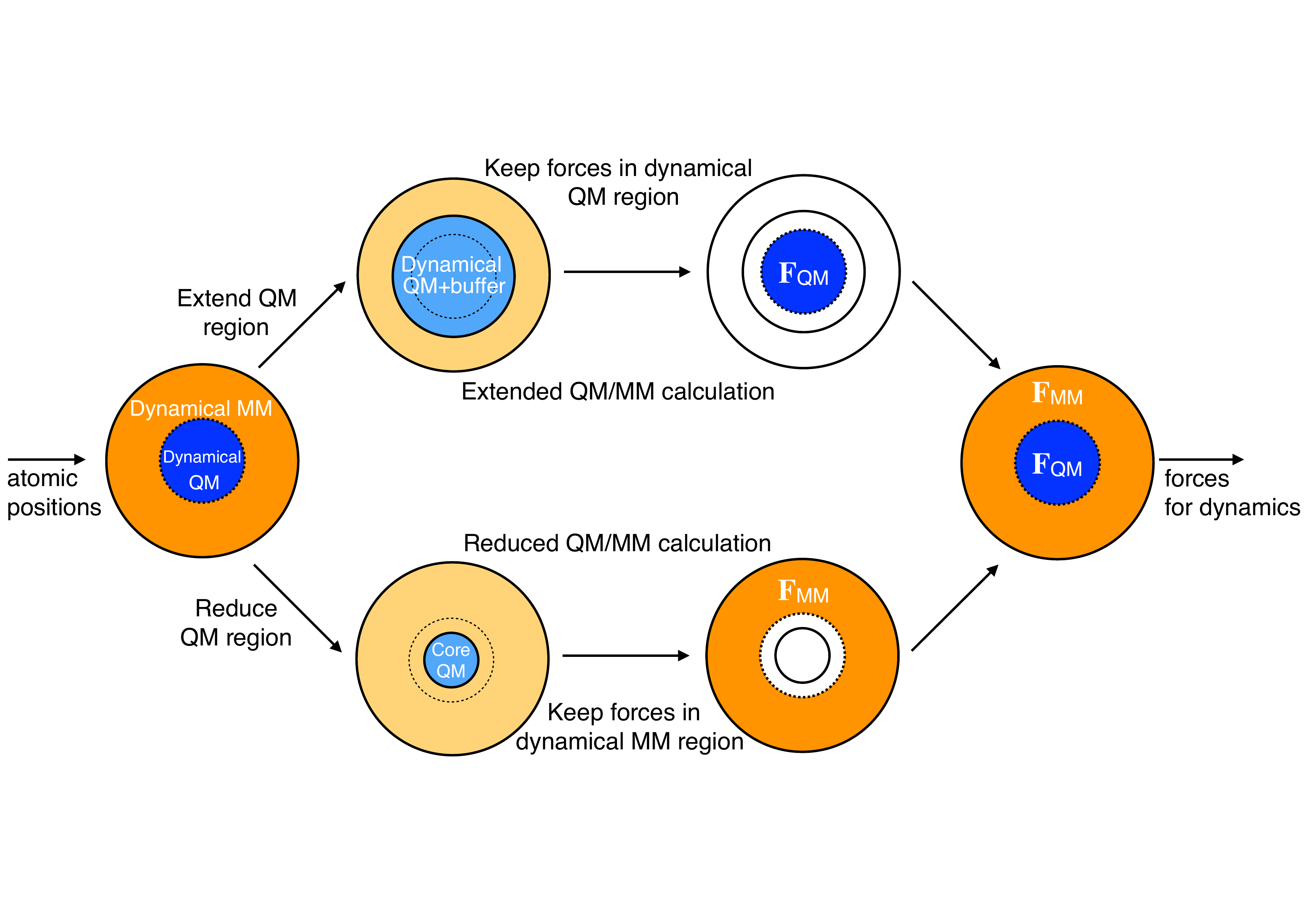}
        \caption{Flowchart of the adaptive buffered force QM/MM
        method. For each configuration during the dynamics two
        calculations are performed: an extended  QM/MM calculation
        to get accurate forces in the dynamical QM region and a
        reduced  QM/MM calculation (that can be fully MM if
        corresponding parameters are available) to get converged
        forces in the dynamical MM region.}
        \label{flowchart}
\end{figure}

In the adbf-QM/MM method the atomic forces that are used in molecular dynamics simulations
to generate a  trajectory are obtained by combining
two QM/MM force calculations.  A flowchart describing the force
calculations is shown in Fig.~\ref{flowchart}. At each time step, the system is partitioned into a number of different regions, which are defined as follows. We begin by creating two sets of atoms, the first consisting of atoms that should follow trajectories using QM forces
(we  call this the \textit{dynamical QM region}), and those that should follow MM forces (\textit{dynamical MM region}).
The first and more expensive QM/MM calculation (``extended
QM/MM calculation'') uses an enlarged QM region to obtain accurate
forces for atoms in the dynamical QM region.  This extended QM region is
constructed by adding a \emph{buffer region} around the dynamical QM region.
The   buffer region size required to reduce the force errors at the QM-MM
boundary  below a preset threshold can be determined from the convergence
of forces in the dynamical QM region as a function of buffer region size, 
carried out separately before the production run on a few  relevant configurations (e.g.\ near the estimated 
extrema of a free energy profile). 

The second QM/MM
calculation (``reduced QM/MM calculation'') uses a smaller QM region
(which we call the \textit{core region}) to reduce force errors due to the
QM-MM boundary on atoms in the MM region.  When the necessary
force field parameters are available, the core  region may be eliminated altogether and
this reduced size  QM/MM calculation replaced by a cheap fully MM calculation.  The forces for the propagation
of the dynamics are then obtained based on the current identity of
the atoms:
\begin{equation}
        \label{eq:force} \mathbf{F}_{i}= \left\{ \begin{array}{l
        l}
                \mathbf{F}_{i}^{\textrm{Extended}}, & \textrm{if}
                \ i \in \textrm{dynamical QM region} \\
                \mathbf{F}_{i}^{\textrm{Reduced}}, & \textrm{if} \
                i \in \textrm{dynamical MM region} \\
        \end{array} \right.
\end{equation}
This is a so-called abrupt force-mixing scheme, where forces used for
dynamics switch from one description to the other without a 
transition region. When an atom is switched from the dynamical QM region to the dynamical MM region or vice versa, the force it experiences  has a discontinuity. Introducing a narrow transition region in which the dynamical force is a linear combination of the forces calculated in the extended and reduced QM/MM calculations would smooth out this discontinuity.\cite{Heyden07,Kerdcharoen96,Kerdcharoen02,Bulo09}  

\begin{figure}
        \centering
        \includegraphics[width=0.5\columnwidth]{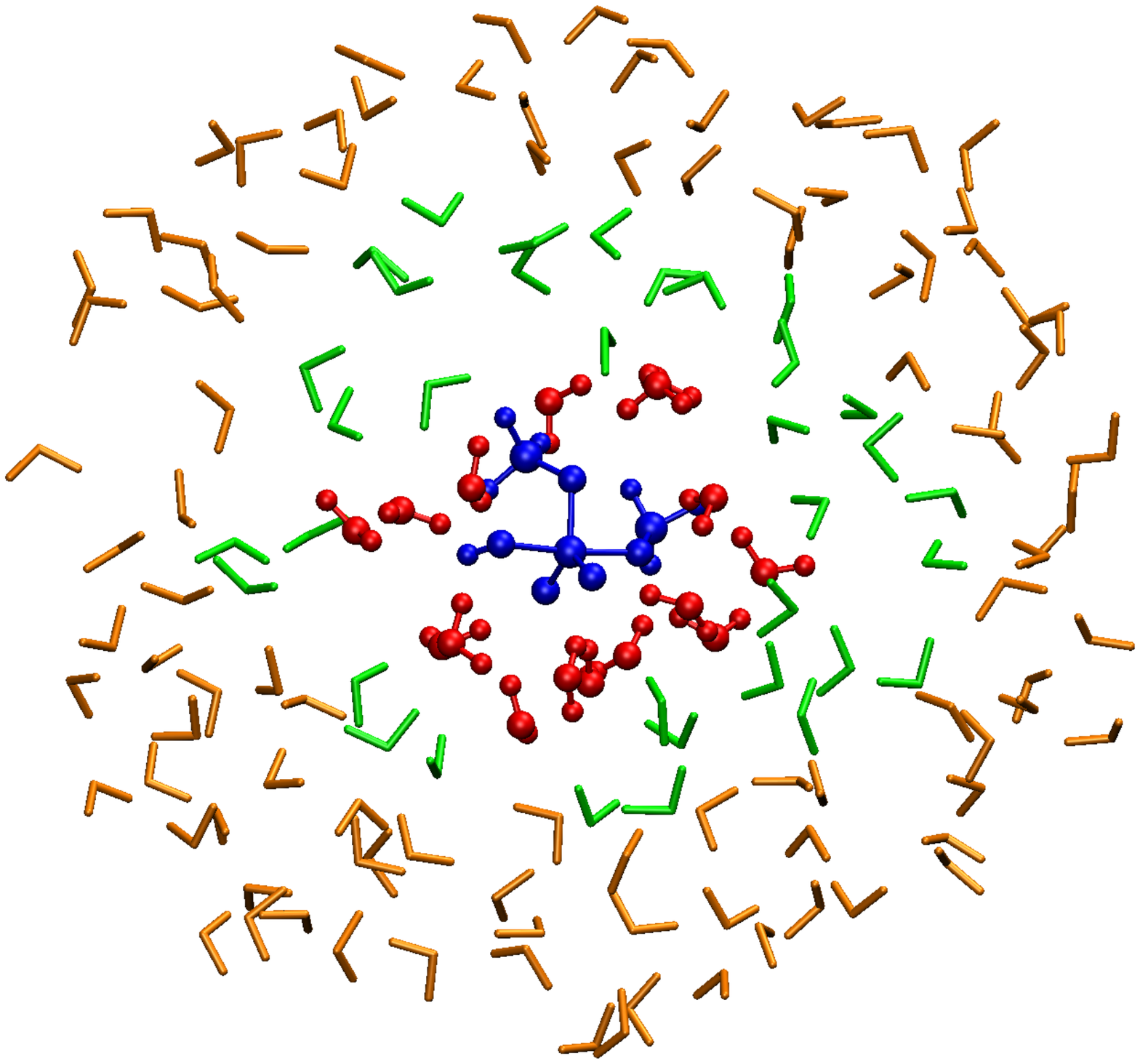}
        \caption{Visualization of the QM regions of an adaptive
        buffered-force QM/MM simulation of dimethyl-phosphate
        hydrolysis. The core region is the dimethyl-phosphate and
        the attacking hydroxide ion (blue) with no additional
        adaptively selected atoms. The dynamical QM region (red)
        is selected by extending the core region by $r_{\mathrm{qm}}
        = 3.0 - 3.5$~{\AA}. The buffer region (green) is an additional
        layer around the dynamical QM region within $r_{\mathrm{buffer}}
        = 3.0 - 3.5$~{\AA}. The rest of the system (orange) is
        treated as MM in both the extended and reduced calculations.
        Ball-and-stick representation is used for atoms which follow QM
        forces in the dynamics.}
        \label{regions}
\end{figure}

Adaptivity is achieved by defining  criteria to select 
atoms for the various regions that are dynamically evaluated at each
time step during  the simulation.   In our implementation each region is composed
of a list of atoms fixed by the user due to their chemical role 
and additional atoms that are selected  due to their distance from atoms in
other regions.
First, the core region is created by combining the fixed list and nearby atoms,
based on a cutoff distance, $r_{\rm core}$,  from the atoms in the fixed list.  Next, the dynamical QM region
is defined as the union of the core region, another (optional) fixed list and atoms within a cutoff distance, $r_{\rm qm}$,  of core
region atoms.  
Finally the buffer region is defined as the union of yet another optional fixed list and atoms
within a  cutoff distance, $r_{\rm buffer}$,   from atoms in the
dynamical QM region.  An example of these regions from a
simulation of the hydrolysis of dimethyl phosphate (at the transition state) is shown in Fig.~\ref{regions}.
To reduce the frequency of switching
between regions for atoms that are close to the boundary, {\em hysteresis} is applied to all distance cutoffs, so an atom has to come 
closer than some inner radius to become incorporated into a region, but must 
move farther than a larger, outer radius to be removed from the region. 

The use of force-mixing has two direct consequences stemming from the lack
of a  total potential energy for the system.
First, because the forces are not the derivatives of any energy  function,
the dynamics are not conservative. Any deviation from linear momentum
conservation is easily fixed exactly by adding a small correction force to
some or all atoms, but the deviation from energy
conservation necessitates the use of an appropriate thermostat to
maintain the correct kinetic temperature throughout the system.  We
have found that a simple adaptive Langevin thermostat\cite{Jones11}
(described below) is sufficient to give a stable and spatially uniform temperature
profile.\cite{Varnai13} Second, the lack of a total energy
prevents the use of some free energy calculation methods, although
potential of mean force methods, which require only forces and
trajectories, can still be applied.\cite{Varnai13}

By appropriately setting the cutoff distances for the various
 regions, the adbf-QM/MM method can be made to be equivalent to a
number of other adaptive methods which we compare to here.  The
adaptive conventional QM/MM method (adconv-QM/MM), which is an energy-mixing scheme and is equivalent
to the ABRUPT method~\cite{Bulo13}, corresponds to setting the  
core and dynamical QM regions to be the same and an empty buffer region.  The
adaptive unbuffered force-mixing QM/MM method (adubf-QM/MM), which
is very close to the hot spot method,~\cite{Kerdcharoen96} corresponds
to an empty (or minimal) core region, an adaptive dynamical QM region, and an empty buffer region.  
The difference between the adconv-QM/MM and adubf-QM/MM methods lies therefore
in how the dynamical forces for the MM atoms are obtained. In the adconv-QM/MM method there is only one QM/MM force calculation, and the MM atoms are propagated using the forces from this same QM/MM force calculation that yields the forces for the QM atoms. In the adubf-QM/MM method, which is a true force-mixing approach, the MM atoms are propagated with forces obtained from either a fully MM calculation or a reduced QM/MM calculation with a very small QM region which includes just the reactants. In addition, we also compare our results to a conv-QM/MM
simulation, which is not adaptive, so only the solutes are treated
quantum
mechanically.
\subsection*{\sffamily \large Implementations of Adaptive Buffered Force QM/MM method}
We have implemented adbf-QM/MM in two popular QM/MM programs:
the AMBER  package\cite{amber}, which has a number of built in
SE methods as well as an interface to external QM
programs, and CP2K, which is primarily a DFT  package
but contains some SE models\cite{cp2k}.  Because of the different
structure of the two codes, the actual implementations are slightly
different, so we begin here with the common and general concepts
needed to specify an adbf-QM/MM calculation.
In addition to the general QM/MM keywords used by each program the user has to specify only a few  additional variables.  The most important ones control the inclusion of atoms in the various regions:
\begin{itemize}

  \item Specification of a disjoint list of fixed core, dynamical QM and
  buffer atoms.  In CP2K the fixed core region  cannot be empty; otherwise these lists are optional. 

  \item Specification of the hysteretic inner ($r_{\textrm{in}}$)
  and outer ($r_{\textrm{out}}$) radii of the adaptive core, dynamical QM
  and buffer regions.  
\end{itemize}

Both the CP2K and AMBER implementations take special care with covalent bonds
crossing the  interfaces in the reduced and extended QM/MM
calculations.  To minimize errors associated with breaking such
covalent bonds indiscriminately, only entire molecules or fragments
bounded by particular covalent bonds are included or excluded from
each region.    In CP2K the specific covalent bonds that can be cut by the
reduced and extended calculations'  interfaces
must be fixed in the input file, and large molecules (such as proteins) 
that should not be entirely included or excluded must therefore be 
omitted from the adaptive region selection.  The AMBER implementation supports an 
adaptive definition of breakable covalent bonds at the  interfaces.

Both implementations support different ways of applying the momentum 
conservation correction. The CP2K implementation supports different
total charges of the QM region in the reduced and extended calculations,
as well as constructing the dynamical QM region based only on distances  from the
fixed subset of the core region.  The AMBER implementation
automatically adjusts the total charge in the reduced and extended
QM/MM calculations based on a default table of oxidation numbers of
the adaptively selected atoms. This table can be modified by the user,
and the AMBER implementation also supports a number of different geometrical criteria for adaptive core,
dynamical QM, and buffer selection.

Adaptive thermostats required for adbf-QM/MM dynamics have been  implemented,
including support for independent thermostats for each degree of freedom, using the adaptive 
Langevin~\cite{Jones11} method (CP2K and AMBER) and several variants of the adaptive 
Nos\'e-Hoover~\cite{Samoletov07, Leimkuhler:2009ge, Jones11}
method (AMBER only).  The adaptive Langevin thermostat is essentially a Langevin thermostat (to ensure ergodicity) in parallel
with a Nos\'e-Hoover thermostat (to compensate for deviations from
energy conservation), and the corresponding dynamical equations are
\begin{eqnarray}
   \dot{q} & = & \frac{p}{m} \\
   \dot{p} & = & F(q) - \left( \gamma + \chi \right) p + \sqrt{2 k_B T \gamma m}\dot{w} \\
   \dot{\chi} & = & \left( 2 K - n k_B T \right) / Q.
\end{eqnarray}
The position and momentum vectors are $q$ and $p$, respectively, $\chi$ is the Nos\'e-Hoover
degree of freedom, $m$ is the atomic mass, and $F(q)$ is the force.  The temperature
is $T$, Boltzmann's constant is $k_B$, $K$ is the kinetic energy, and $n$
is the number of degrees of freedom associated with the thermostat.
The Langevin friction is $\gamma = 1/\tau_L$ where $\tau_L$ is the Langevin
time constant, the Nos\'e-Hoover fictitious mass is $Q=k_B T \tau_\mathrm{NH}^2$ where
$\tau_\mathrm{NH}$ is the Nos\'e-Hoover time constant, and $\dot{w}$
is the time derivative of a Wiener process.  The adaptive Nos\'e-Hoover method has a similar
structure, but the Langevin thermostat is replaced with Nos\'e-Hoover
chains with an optional Langevin thermalization of the last thermostat
in the chain.  In its most general form this gives the adaptive Nos\'e-Hoover-chains-Langevin
method with the corresponding equations
\begin{eqnarray}
   \dot{q} & = & \frac{p}{m} \\
   \dot{p} & = & F(q) - \left( \xi_{1} + \chi \right) p \\
   \dot{\xi_{1}} & = & \left( 2K - n k_B T \right) / Q_{1} - \xi_{1} \xi_{2} \\
   \dot{\xi_{2}} & = & \left( Q_{1}\xi_{1}^{2} - k_BT \right) / Q_{2} - \xi_{2} \xi_{3} \\
   \dots \nonumber \\
   \dot{\xi_{r}} & = & \left( Q_{r-1}\xi_{r-1}^{2} - k_BT \right) / Q_{r} + \sqrt{2 k_B T \gamma_l Q_r} \dot{w} - \gamma_l \xi_r \\
   \dot{\chi} & = & \left( 2 K - n k_B T \right) / Q,
\end{eqnarray}
where $r$ is the length of the chain, $\xi_i$ and $Q_i$ are the Nos\'e-Hoover
chain degrees of freedom and their masses, respectively, and $\gamma_l$ is the Langevin friction
for thermalizing the final thermostat in the chain.  
Setting $r$ to 1 corresponds to the adaptive Nos\'e-Hoover-Langevin thermostat, while omitting the 
Langevin part (i.e.\ formally setting $\gamma_l$ to 0) with $r>1$ results in the adaptive Nos\'e-Hoover-chain.

Both adaptive thermostats can be applied so that a separate  NH variable (or NH chain)\
 is coupled to each degree of freedom \cite{Tuckerman93}, rather than a single NH variable coupling to the total kinetic energy.
This is the mode in which we use adaptive thermostats in this work,
because in the nonconservative force-mixing simulations extra heat
is generated locally near the QM-MM interface and the amount that
needs to be dissipated therefore varies in space.

\begin{figure}
        \centering
        \begin{subfigure}{1.0\columnwidth}
                \includegraphics[width=0.5\columnwidth]{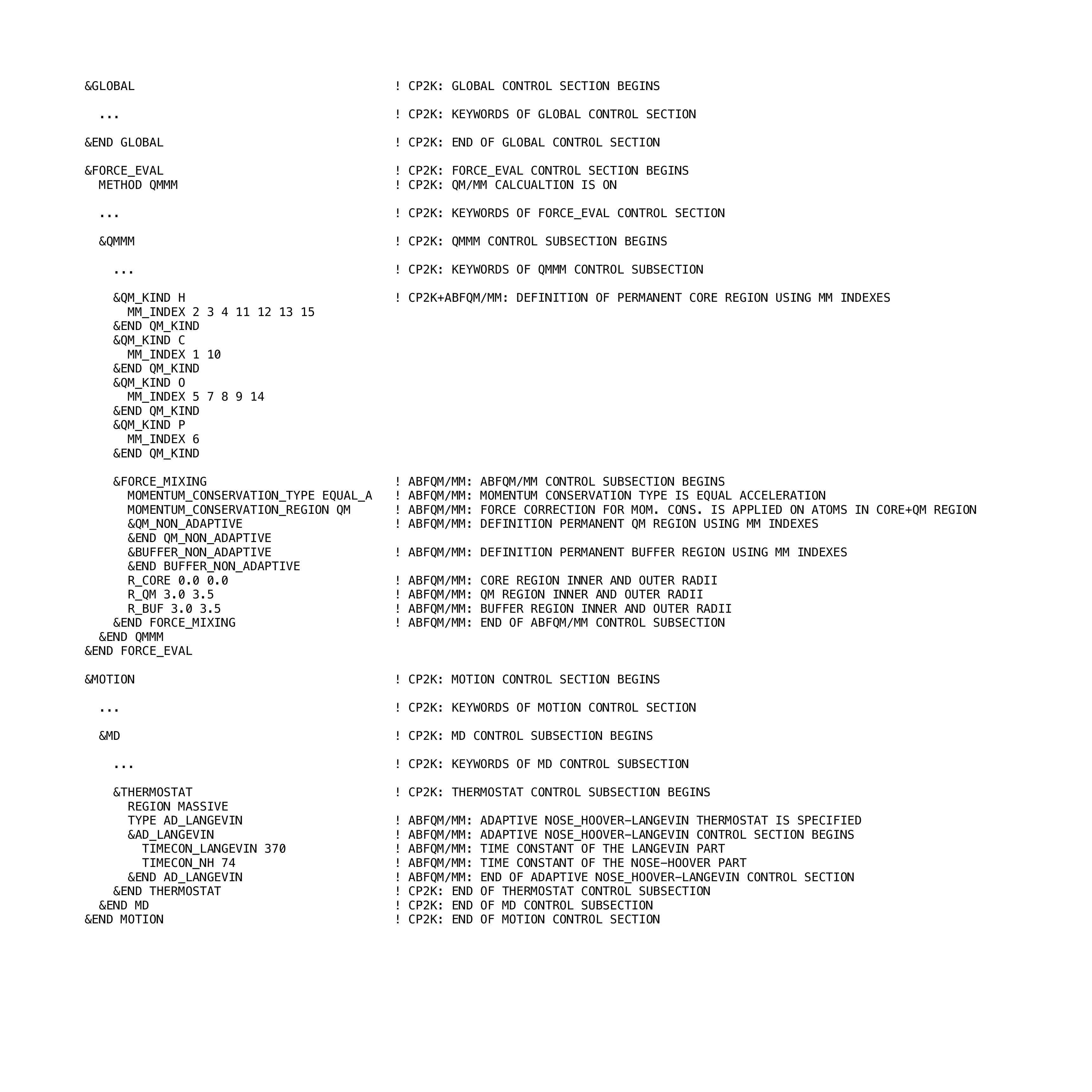}
                \label{fig:input_cp2k}
                \caption{}
        \end{subfigure}
        \begin{subfigure}{1.0\columnwidth}
                \includegraphics[width=0.5\columnwidth]{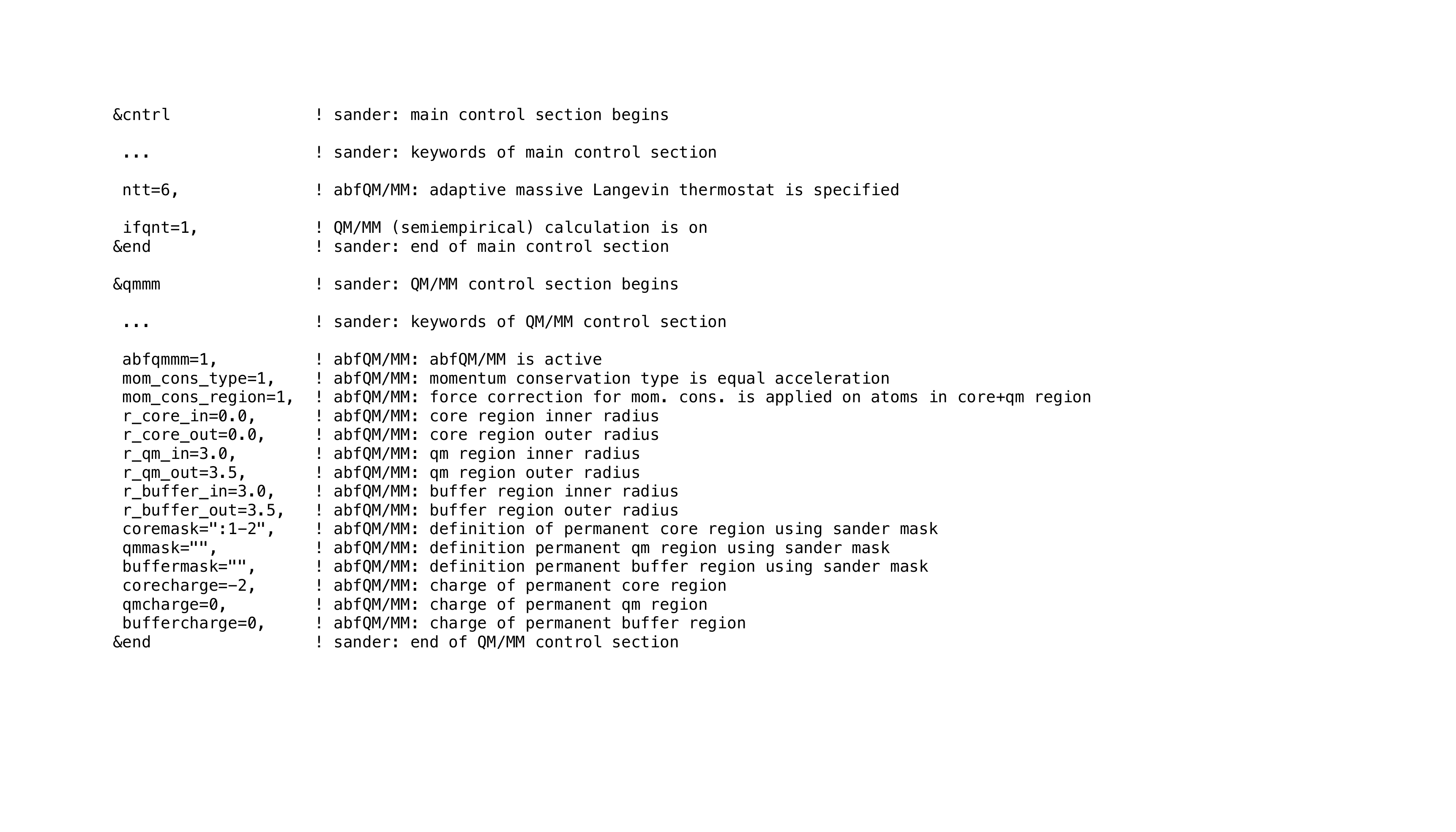}
                \label{fig:input_amber}
                \caption{}
        \end{subfigure}
        \caption{Basic CP2K (a) and AMBER (b) input files extended by the adbf-QM/MM related keywords of the phosphate system.}
        \label{fig:input}
\end{figure}

The sections of example CP2K and AMBER input files that are relevant
to the adbf-QM/MM implementations are shown in Fig.~\ref{fig:input}.
The CP2K inputs consist of a  {\tt \&QMMM} section to
specify the fixed core list, a {\tt \&FORCE\_MIXING} section to specify
the other regions and momentum conservation details, and a {\tt
\&THERMOSTAT} section with a {\tt REGION MASSIVE} keyword and an
{\tt \&AD\_LANGEVIN} section specifying the two time constants.
The AMBER input specifies the thermostat with the {\tt ntt} keyword (6, 7 and 8 for the adaptive Langevin, adaptive Nos\'e-Hoover chains and adaptive Nos\'e-Hoover chains with Langevin, respectively),
activates the QM/MM functionality, and enables force-mixing in
the {\tt \&qmmm} section with the {\tt abfqmmm=1} keyword.  In this
section the input file also sets the momentum conservation details, fixed
lists and adaptive core, dynamical QM, and buffer radii, as well as the charges
of the three regions.  Example   input files are included as
supplementary information, however  these  do not show every available option,
and full details are available in the documentations of the two
packages.  

\subsection*{\sffamily \large Model Systems}

To test the adaptive QM/MM implementations we studied  structure and
reaction free energy profiles in three systems.  In pure bulk water,
which provides a stringent test for adaptive
methods as previous work has shown,\cite{Bernstein12} we studied the structure for a number
of QM models and adaptive QM/MM methods.  For two reactions in water solution,
the autoprotolysis of water in the presence of a Zn$^{2+}$ ion and the hydrolysis
of dimethyl-phosphate, we calculated the free energy profile using a
number of adaptive QM/MM methods.  In all cases we compared to  reference calculations
employing a fully QM description using smaller simulation cells, and for the autoprotolysis
of water we also ran  fully QM simulations using an intermediate size unit cell.
The QM region sizes for all QM/MM simulations are summarized in Table~\ref{tab:radii}.
Adaptive radii were applied to distances between all atoms, except for SE bulk water simulations
where only O-O distances were used to select molecules.
The sum of core and dynamical QM radii were chosen to ensure that the first hydration shell is included in the dynamical QM region. 
\begin{table}
\centering
\begin{tabular}{l c c c}
\toprule
Simulation type & $r_{\mathrm{core}}$ (\r{A}) & $r_{\mathrm{qm}}$ (\r{A}) & $r_{\mathrm{buffer}}$ (\r{A}) \\
\midrule
\emph{SE Bulk water} \\
adbf-QM/MM & 0.0 -- 0.0 & 4.0 -- 4.5 (*) & 4.0 -- 4.5 (*)\\
\midrule
\emph{MNDOd Autoprotolysis reaction} \\
conv-QM/MM & 0.0 -- 0.0 & 0.0 -- 0.0 & 0.0 -- 0.0 \\
adconv-QM/MM & 2.5 -- 3.0 & 0.0 -- 0.0 & 0.0 -- 0.0 \\
adubf-QM/MM & 0.0 -- 0.0 & 2.5 -- 3.0 & 0.0 -- 0.0 \\
adbf-QM/MM & 0.0 -- 0.0 & 2.5 -- 3.0 & 3.0 -- 3.5 \\
\midrule
\emph{DFT bulk water and dimethyl-phosphate hydrolysis} \\
conv-QM/MM & 0.0 -- 0.0 & 0.0 -- 0.0 & 0.0 -- 0.0 \\
adconv-QM/MM & 3.0 -- 3.5 & 0.0 -- 0.0 & 0.0 -- 0.0 \\
adubf-QM/MM & 0.0 -- 0.0 & 3.0 -- 3.5 & 0.0 -- 0.0 \\
adbf-QM/MM & 0.0 -- 0.0 & 3.0 -- 3.5 & 3.0 -- 3.5 \\
\bottomrule
\end{tabular}
\caption{Adaptive region radii for the QM/MM simulations, applied to all interatomic distances, except
for SE bulk water simulations (*), where the selection criterion was based only on the oxygen--oxygen distances.}
\label{tab:radii}
\end{table}

All systems were simulated using constant temperature and volume
molecular dynamics. For bulk water the structure was analyzed by
calculating the time averaged radial distribution function (RDF) for a
molecule at the center of the dynamical QM region.  Free energy profiles were
calculated using umbrella integration (UI)\cite{Kastner05}, with a bias
potential 
$$
V_{\mathrm{restraint}} = \frac{1}{2} k\left(x(\mathbf{r}) -x_{0}\right)^{2}
$$
where  $k$ is the curvature,  $x_0$ is the desired value of the collective coordinate, and
$x(\mathbf{r})$ is its instantaneous value.
In the biased simulation the mean gradient of the bias potential
is approximately equal to the negative of the gradient of the
potential of mean force (PMF) at the mean value of the collective
coordinate\cite{Kastner05}. For simulations with AMBER the bias was achieved using the PMFlib  package\cite{pmflib}
that was linked to AMBER, and for CP2K internal subroutines
were used.

\emph{Bulk water structure}

For bulk water we used cubic simulation cells with 13.8~{\AA} (93 molecules)
and 41.9~{\AA} (2539 molecules) sides for the fully QM and QM/MM 
calculations, respectively.  The MM water molecules were described
with the flexible TIP3P (fTIP3P) potential.~\cite{MacKerell98}  We used the
AMBER implementation to compare the results of the adbf-QM/MM method
for a number of SE models.  In each simulation a single water
molecule was selected to be the center of the dynamical QM region, with
radii listed in Table~\ref{tab:radii} applied only to O-O distances 
when selecting molecules for the adaptive regions.  No core region
was used, so the reduced size calculation was done as a fully MM
calculation.  The SE models compared were MNDO\cite{Dewar77},
AM1\cite{Dewar85}, AM1d\cite{Nam07}, AM1disp\cite{Korth10},
PM3\cite{Stewart89}, PM3-MAIS\cite{Uruchurtu00}, PM6\cite{Stewart07},
RM1,\cite{Rocha06} and DFTB\cite{Porezag95}. Using the CP2K
implementation we compared the results of various QM/MM methods\cite{Laino05,Laino06}
with DFT and the BLYP exchange-correlation functional\cite{Becke88,
Lee88, Miehlich89} plus Grimme's van der Waals correction,\cite{Grimme06,Wang11}
with a DZVP basis, GTH pseudopotentials,~\cite{Goedecker96} and a density cutoff of 280~Ry.
The methods compared were conv-QM/MM, adubf-QM/MM, adconv-QM/MM,
and adbf-QM/MM.  In this case a single water molecule was selected for the fixed
core region, with adaptive radii listed in Table~\ref{tab:radii}
applied to all interatomic distances.  

\emph{Reaction free energy profiles}

\begin{figure}
        \centering
        \includegraphics[width=0.5\columnwidth, trim=2 2 2 2, clip]{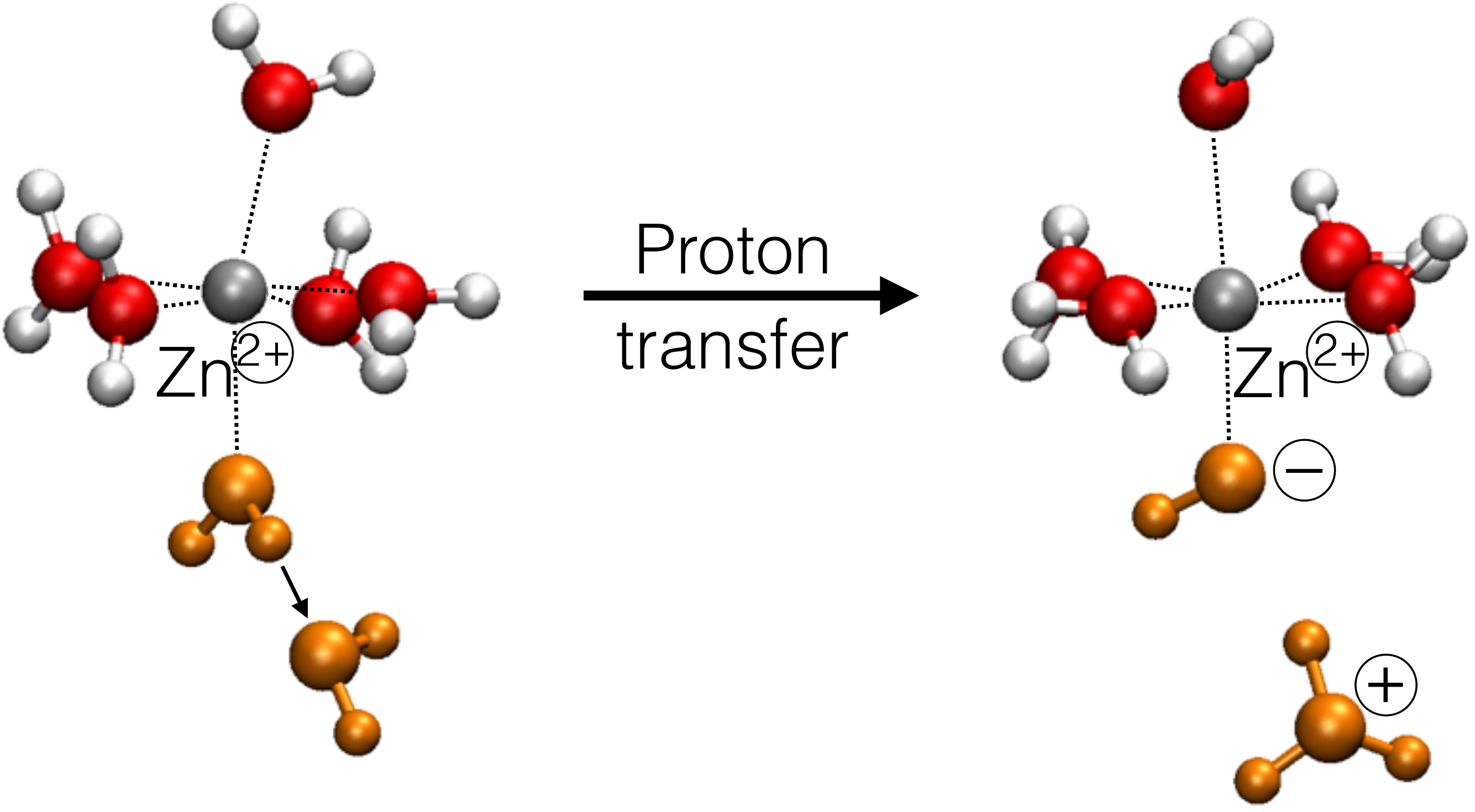}
        \caption{Reaction scheme of the autoprotolysis between a zinc-coordinated and a non-coordinated water molecules (orange).}
        \label{fig:auto}
\end{figure}

Water related proton transfer reactions can be facilitated by the
presence of divalent metal ions\cite{Jencks87}. The metal ion
lowers the \emph{p}K$_{a}$ of the coordinated water molecule making it
a stronger acid. Our example is a very simple model of this
phenomenon, the proton transfer reaction
between a zinc-coordinated water molecule (proton donor) and a
non-coordinated water molecule (proton acceptor) in water solution, 
shown in Fig.~\ref{fig:auto}.
To calculate the free energy profile for this reaction we used
UI with   the collective coordinate being the difference between 
rational coordination numbers (\emph{DRCN}) of the acceptor and donor oxygen atoms\cite{Sprik98, Mones12}:
\begin{equation}
\label{eq:drcn}
\text{\emph{DRCN}}(\left\{\mathbf{r}_{\text{HO}_{D}}, \mathbf{r}_{\text{HO}_{A}}\right\})=\text{\emph{RCN}}(\left\{\mathbf{r}_{\text{HO}_{A}}\right\})-\text{\emph{RCN}}(\left\{\mathbf{r}_{\text{HO}_{D}}\right\})
\end{equation}
and
\begin{equation}
\label{eq:rcn}
\text{\emph{RCN}}\left(\left\{\mathbf{r}_{\text{HO}_{D/A}}\right\}\right)=\sum^{\mathrm{H \ atoms}}_{i}\frac{1-(\frac{r_{i}}{r_{0}})^\alpha}{1-(\frac{r_{i}}{r_{0}})^\beta},
\end{equation} 
where the subscripts \emph{D} and \emph{A} denote the donor and acceptor oxygen atoms, respectively, $\alpha=6$, $\beta=18$ and the reference distance $r_{0}=1.6$~{\AA}.

The reactions were simulated in cubic cells with sides of
13.6~{\AA} (87 water molecules) and 17.2~{\AA} (174 water molecules)
for the fully QM and 45.8~{\AA} (3303 water molecules)
for the QM/MM simulations.  The simulations
were carried out using the AMBER implementation with Zn$^{2+}$ ion parameters
from Ref.~\cite{Hoops91} , fTIP3P model for MM waters,\cite{MacKerell98} and the MNDO(d)
SE method.~\cite{Thiel92}  The Zn$^{2+}$ ion and two reactant water
molecules were defined as the QM region in the conv-QM/MM simulation,
as well as the fixed core region in the adaptive simulations.
Adaptive regions used radii listed in Table~\ref{tab:radii} with
all interatomic distances and only entire water molecules included
or excluded in any region.

In all autoprotolysis simulations we applied one-sided harmonic restraints for the following
3 distances: one between the two O atoms beyond 3.0~{\AA} to
keep the reactants together, another between the O atom
of donor water molecule and zinc ion beyond 2.5~{\AA} to keep the
donor water molecule in the coordination sphere of the metal ion,
and the third between the O atom of acceptor water molecule
and zinc ion for distances larger than 3.5~{\AA} to prevent the
acceptor water molecule from entering into the coordination sphere
of the metal ion. For each restraint a force constant of
25.0~kcal~mol$^{-1}$~{\AA}$^{-2}$ was applied.  The applied force
constant for the UI restraint was 400~kcal~mol$^{-1}$ and the profile
was calculated in the range of \emph{DRCN}~$\in [-0.2, 2.2]$.

\begin{figure}
        \centering
        \includegraphics[width=0.5\columnwidth, trim=2 2 2 2, clip]{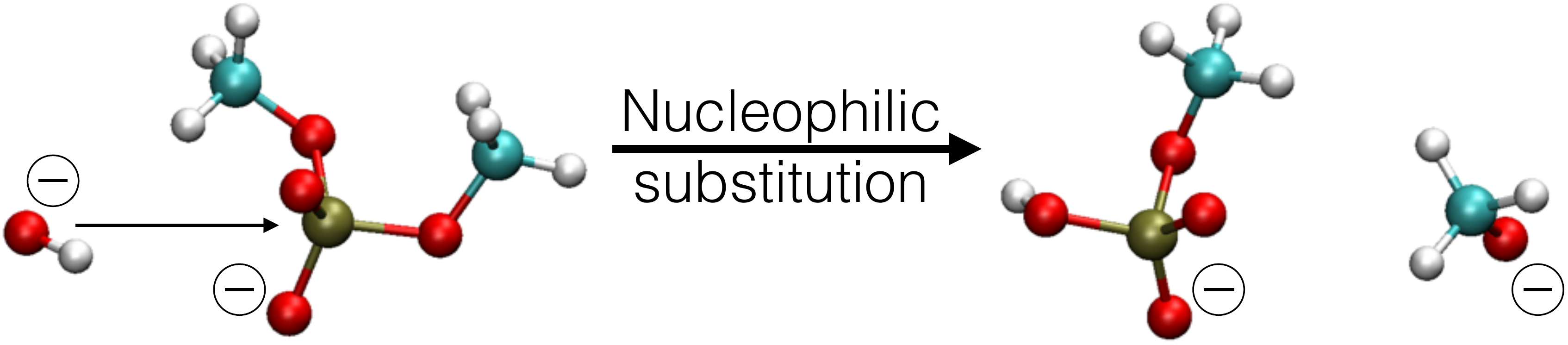}
        \caption{Reaction scheme of the dimethyl-phosphate hydrolysis.}
        \label{fig:dmpoh}
\end{figure}

The second reaction we simulated was dimethyl-phosphate hydrolysis,
shown in Fig.~\ref{fig:dmpoh}, where an incoming hydroxide ion
attacks the dimethyl-phosphate and causes a methoxide ion
to leave. A similar hydrolysis of phosphate diesters in solution is a biologically important
type of phosphoryl transfer reactions and a key model to understand
DNA cleavage~\cite{Kamerlin13}.
The reaction coordinate for the UI procedure was the distance difference 
between the leaving O-P atoms and the attacking O-P atoms
\begin{equation}
\label{eq:dd}
\text{\emph{DD}}(\mathbf{r}_{\text{PO}_{L}}, \mathbf{r}_{\text{PO}_{A}})=\left|\mathbf{r}_{\text{PO}_{L}}\right|-\left|\mathbf{r}_{\text{PO}_{A}}\right|,
\end{equation}
where \emph{L} and \emph{A} designate the leaving and attacking O atoms, respectively. 
The reaction was simulated in cubic cells with sides of 13.6~{\AA}
(86 water molecules) and 48.4~{\AA} (3903 water molecules) for the fully
QM and QM/MM simulations, respectively.

Because our simulation protocol starts with an  MM relaxation, MM parameters were needed for the solutes.   The charges of the phosphate
and hydroxide were calculated according to the standard
procedure\cite{Bayly93, Cornell93}, while the bonded and vdW
parameters of the phosphate were derived from the ff99SB version of the AMBER force
field\cite{Hornak06}, and the water molecules were described by the
fTIP3P model.~\cite{MacKerell98}  For the hydroxide ion the same parameters
were used as for the fTIP3P.  For
the DFT model the BLYP exchange-correlation functional\cite{Becke88,
Lee88, Miehlich89} was applied with Grimme's van der Waals correction,\cite{Grimme06,Wang11}
using the DZVP basis set with GTH
pseudopotentials\cite{Goedecker96} and a density cutoff of 280~Ry.
The QM region of the conv-QM/MM calculation and the fixed core
region of the adaptive QM/MM calculations consisted only of the
reactant dimethyl-phosphate and hydroxide.  Adaptive regions used
the radii listed in Table~\ref{tab:radii} with all interatomic
distances and only entire water molecules selected for inclusion
or exclusion.  The free energy profile was carried out in the range
of \emph{DD}~$\in [-3.0, 3.0]$~{\AA} using an UI restraint force constant of 
400~kcal~mol$^{-1}$~{\AA}$^{-1}$.

\subsection*{\sffamily \large Simulation protocol}
\label{sec:methods:protocol}

\emph{General simulation parameters}

All simulations used periodic boundary conditions with MM-MM
electrostatic interactions calculated by the Ewald\cite{Ewald21}
and particle-mesh Ewald\cite{Darden93} for the small and large
simulation cells, respectively.  For fully SE  and DFT simulations,
the CP2K package was used with the smooth particle mesh Ewald method and
multipole expansion up to quadrupoles.~\cite{Laino03}  In the AMBER
QM/MM simulations the QM-MM interactions were calculated using a 
multipole description within 9~{\AA} while both the long-range QM-QM
and QM-MM electrostatic interactions were based on the Mulliken
charges of the QM atoms according to Ref.\cite{Nam05, Saebra07, Walker08, Gotz14}.  In the CP2K
QM/MM simulations the QM-MM interaction used Gaussian smearing of
the MM charges.~\cite{Laino05} When systems were charged a uniform
background countercharge was applied.
Molecular dynamics simulations with a time step of 0.5~fs were used for equilibration and
canonical ensemble sampling.

The first step in the simulation protocol was to generate independent initial configurations
for all box sizes from long equilibrium fully MM simulations. In the case of bulk water
all fully QM and QM/MM simulations were started from these MM equilibrated configurations.
For the reactions, first the relatively computationally inexpensive conv-QM/MM
simulations were carried out starting from an initial configuration that was taken from a fully MM   equilibrium simulation at the initial  restraint position corresponding to the reactant state.  The restraint forces
for UI were sampled for some time period, and the restraint center
was slowly changed to the next collective coordinate value, then the process
repeated until the desired range of values were sampled.  The more computationally
expensive fully QM and adaptive QM/MM simulations were started from the final
configuration of each conv-QM/MM trajectory at each restraint center position.

\emph{Initial configurations}

The systems and topologies for investigating the bulk water were created by the Leap program
of the AMBER package~\cite{amber}. The initial geometries
were relaxed for 5000 minimization steps, followed by a molecular
dynamics \emph{NVT} simulation of heating from $T=0$~K to $T=300$~K over 50~ps
followed by 50~ps at fixed temperature.  The density was then relaxed
by a 200~ps \emph{NpT} simulation at $T=300$~K and $p=1$~bar, and then the average box size
was calculated during an additional 500~ps long \emph{NpT} simulation.
During this last stage 10 independent configurations were selected
at 50~ps intervals, which were all rescaled to the mean volume.
Finally, for each of the 10 configurations a 500~ps long \emph{NVT} simulation was
carried out at 300~K. In each case the temperature was controlled by a Langevin 
thermostat\cite{Adelman76} with a friction coefficient of 5~ps$^{-1}$.
The systems for the reactions were also generated using the
Leap program of the AMBER package~\cite{amber} to surround the
reactants by water molecules. These starting configurations were equilibrated
by the same procedure as for the bulk water systems.    

\begin{table}
\centering
\begin{tabular}{l c c c}
\toprule
\multirow{2}{*}{Simulation type} & \multirow{1}{*}{\# of independent} & \multicolumn{2}{c}{Trajectory length per config.} \\
& configurations &  total (ps) & used for analysis (ps) \\
\midrule
\emph{Bulk water} \\
SE fully QM & 10 & 10 & 5 \\
SE adbf-QM/MM & 10 & 50 & 40 \\ 
DFT fully QM & 5 & 10 & 9 \\
\midrule
\emph{Autoprotolysis reaction} \\
MNDOd fully QM & 10 & 12 & 10 \\
MNDOd conv-QM/MM & 10 & 10 & 8 \\
MNDOd adconv-QM/MM & 10 & 5.5 & 4.5 \\
MNDOd adubf-QM/MM & 10 & 5.5 & 4.5 \\
MNDOd adbf-QM/MM & 10 & 5.5 & 4.5 \\
\midrule
\emph{Dimethyl-phosphate hydrolysis reaction} \\
DFT fully QM & 5 & 5 & 2.5 \\
\bottomrule
\end{tabular}
\caption{Configuration numbers and trajectory lengths}
\label{tab:simulations}
\end{table}

\emph{Water autoprotolysis}

Conventional QM/MM simulations were carried out using AMBER and
PMFlib for \emph{DRCN} from -0.2 to 2.2 in increments of 0.1.  The
restraint reaction coordinate was changed from its actual value in the 
reactant state to the starting value of -0.2 over
20~ps.  Then, the \emph{DRCN} was sequentially changed by 0.1 over
1~ps long trajectories, followed by simulation at fixed restraint
position. Restraint force values for UI were collected for the
number of initial configurations and trajectory lengths listed
in Table~\ref{tab:simulations}.  All simulations used a Langevin
thermostat\cite{Adelman76} with a friction coefficient of 5~ps$^{-1}$.

Fully QM simulations for both box sizes were carried out using CP2K, starting from
relaxed conv-QM/MM configurations at each reaction coordinate value,
with a number of independent initial configurations and trajectory
lengths listed in Table~\ref{tab:simulations}.  Temperature was
controlled by the CSVR thermostat\cite{Bussi07} with a time constant
of 200~fs.
Adaptive QM/MM simulations were carried out starting from relaxed
conv-QM/MM for the number of initial configurations and trajectory
lengths listed in Table~\ref{tab:simulations}.  Because of the
energy conservation violation of all the adaptive methods, temperature
was controlled by adaptive Langevin
thermostats~\cite{Jones11}, one per degree of freedom, with a Langevin time constant of 200 fs
and a Nos\'e-Hoover time constant of 200 fs.

\emph{Dimethyl-phosphate hydrolysis}

Initial conditions for the DFT simulations were generated by a
conv-QM/MM simulation with the AM1 SE method using the AMBER code
for \emph{DD} from -3.0~{\AA} to 3.0~{\AA}.  The \emph{DD} was
changed from its initial value to -2.0~{\AA} over 20~ps.  The
\emph{DD} was then changed by increments of 0.1~{\AA} over 1~ps,
followed by equilibration for 10~ps at each \emph{DD} value.  All
subsequent simulations were carried out using CP2K using one
adaptive Langevin thermostat per degree of freedom with a Langevin time constant of 300~fs
and a Nos\'e-Hoover time constant of 74~fs.  Simulations with fully QM,
conv-QM/MM, adconv-QM/MM, adubf-QM/MM, and adbf-QM/MM were carried
out with the number of configurations and trajectory lengths listed
in Table~\ref{tab:simulations}.  Values of \emph{DD} from -3.0~{\AA}
in increments of 0.6~{\AA}, with additional samples at \emph{DD}$=\pm 0.3$~{\AA}
and \emph{DD}$=\pm 0.1$~{\AA}, were used to calculate the UI free energy
profile.

\section*{\sffamily \Large RESULTS}

\subsection*{\sffamily \large Bulk water}

\begin{figure}
        \centering
        \includegraphics[width=0.5\columnwidth]{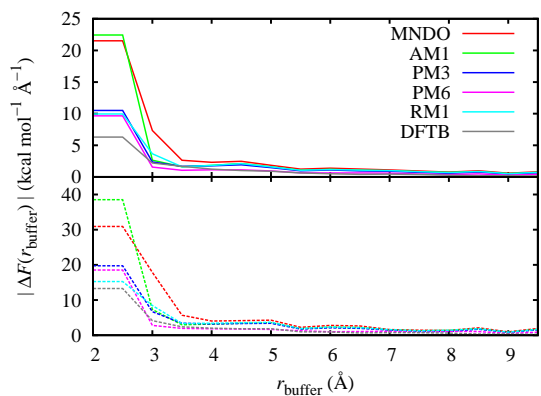}
        \caption{Force convergence on the central oxygen atom in pure bulk water for different SE methods relative to reference forces from calculations using the same SE method with buffer size of 10.0~\AA.
        Top panel shows the mean force error based on 10 independent configurations, and bottom panel shows the maximum error.}
        \label{fig:bulk_force_conv}
\end{figure}

We
performed a force convergence test to determine the appropriate buffer radii by calculating the forces on an
O atom in the center of the QM region of a conventional QM/MM calculation,
as a function of QM region radius, using a number of SE methods.
Here the radius of the QM region models  $r_\mathrm{buffer}$
in the adbf-QM/MM method's extended QM calculation, since it controls
the distance between the molecule whose forces we are testing and
the QM-MM interface.  The atomic configurations were taken from the
10 MM equilibrated configurations described in
the Simulation Protocol subsection
and the calculations were carried out
with MNDO, AM1, PM3, PM6, RM1, and DFTB.  The resulting force errors
calculated with respect to reference forces from a 10~{\AA} radius
conventional QM/MM calculation are plotted in Fig.~\ref{fig:bulk_force_conv}.
For each QM method a similar behaviour is seen in the force
convergence: the average force error goes below
2~kcal~mol$^{-1}$~{\AA}$^{-1}$ (and the maximum goes below 4~kcal~mol$^{-1}$~{\AA}$^{-1}$)
around $r_\mathrm{buffer} = 4.0$~{\AA}, which was chosen as the
lower limit of the buffer size for the dynamics. A similar behaviour was observed in the case of DFT
(BLYP).\cite{Bernstein12}  We also investigated forces on the hydrogen 
atoms (data not shown) and found a slightly faster convergence.

\begin{figure}
        \centering
        \includegraphics[width=0.5\columnwidth]{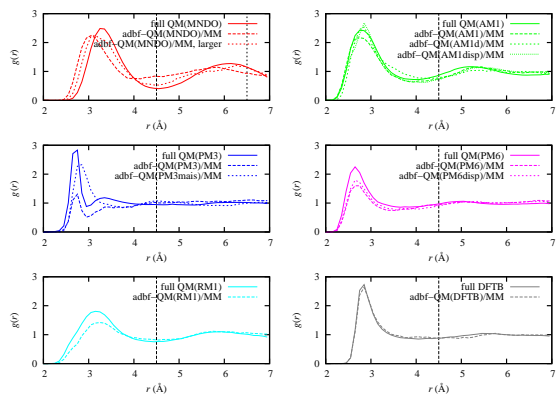}
        \caption{Oxygen--oxygen RDFs in bulk water using different SE methods. Vertical dashed lines at 4.5~{\AA} denote the size of dynamical QM region. For MNDO, a second vertical line at 6.5~{\AA} represents the outer boundary of the dynamical QM region for the larger
simulation.}
        \label{fig:bulk_rdf}
\end{figure}

The oxygen--oxygen RDF averaged over 
10 independent trajectories are plotted in Fig.~\ref{fig:bulk_rdf}.  In the
case of PM3 the fluid density gradually goes down in the dynamical QM region  during the dynamics and
longer simulations showed that this process is irreversible, leading
to an almost complete depletion of water in the dynamical QM region. This phenomenon
was previously observed in Ref.\cite{Winfield09} and the significantly different
diffusion behaviours of the QM and MM water molecules were suggested
as a possible reason. 
The PM3-MAIS method, which is an extension to PM3 parametrised to accurately
reproduce the intermolecular interaction potential of water, does not suffer from this problem. 
In contrast to PM3, for MNDO the water structure in
the QM region is stable for the duration of our simulations but the 
RDF slightly differs from the fully QM result. As expected, using a larger
QM region improves the structure in this case. We also note that the
force convergence for the MNDO is the slowest among the examined
potentials (Fig.~\ref{fig:bulk_force_conv}), so a larger buffer region may 
further improve the RDF.  In the case of PM6 and RM1, the adbf-QM/MM RDFs
show a somewhat lower first peak compared to the fully QM structure.
However, the RDFs remain stable for longer simulation times.
Based on our data we are not able to exclude unambiguously the possibility
that, similarly to PM3, a net flux of atoms leaving the dynamical QM
region causes this discrepancy.  Even if this is the case, the diffusion
is much slower than for PM3.  For DFTB and AM1 the adbf-QM/MM and
fully QM RDFs match almost perfectly.  We investigated two additional
AM1 variants (AM1d and AM1disp) and found similar RDFs to the
fully QM AM1 result.  In general we see that the first peak is
higher for the fully QM simulations than those of adbf-QM/MM.
Although using larger
dynamical QM and buffer regions could potentially improve the agreement,  the improvement
may be limited by differences in how the long range interactions
are calculated~\cite{Murdachaew11,Nam05, Walker08} in the fully QM and 
the adbf-QM/MM simulations due to limitations in the  packages used (CP2K and AMBER, respectively).

\begin{figure}
        \centering
                \includegraphics[width=0.5\columnwidth]{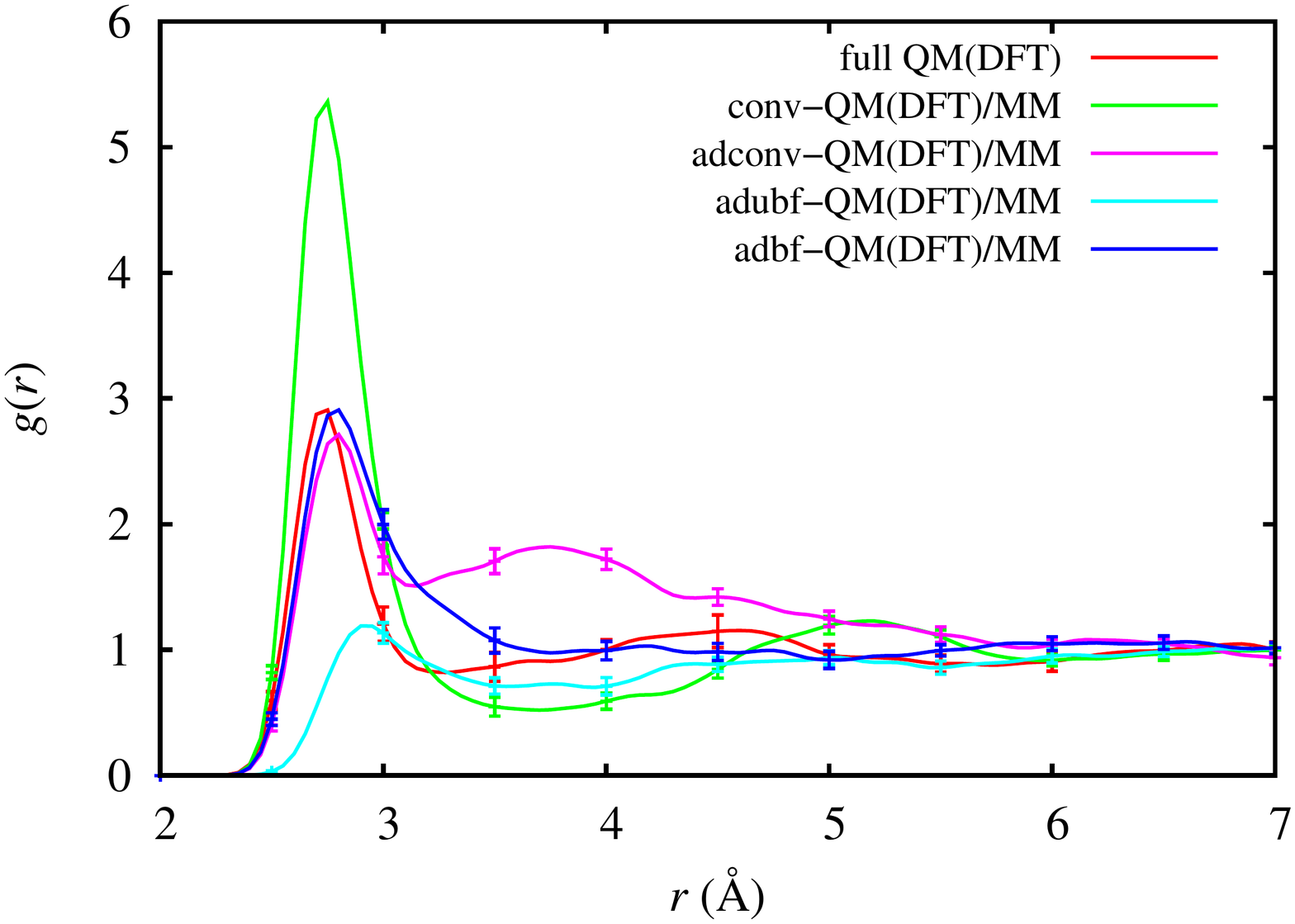}
                \includegraphics[width=0.5\columnwidth]{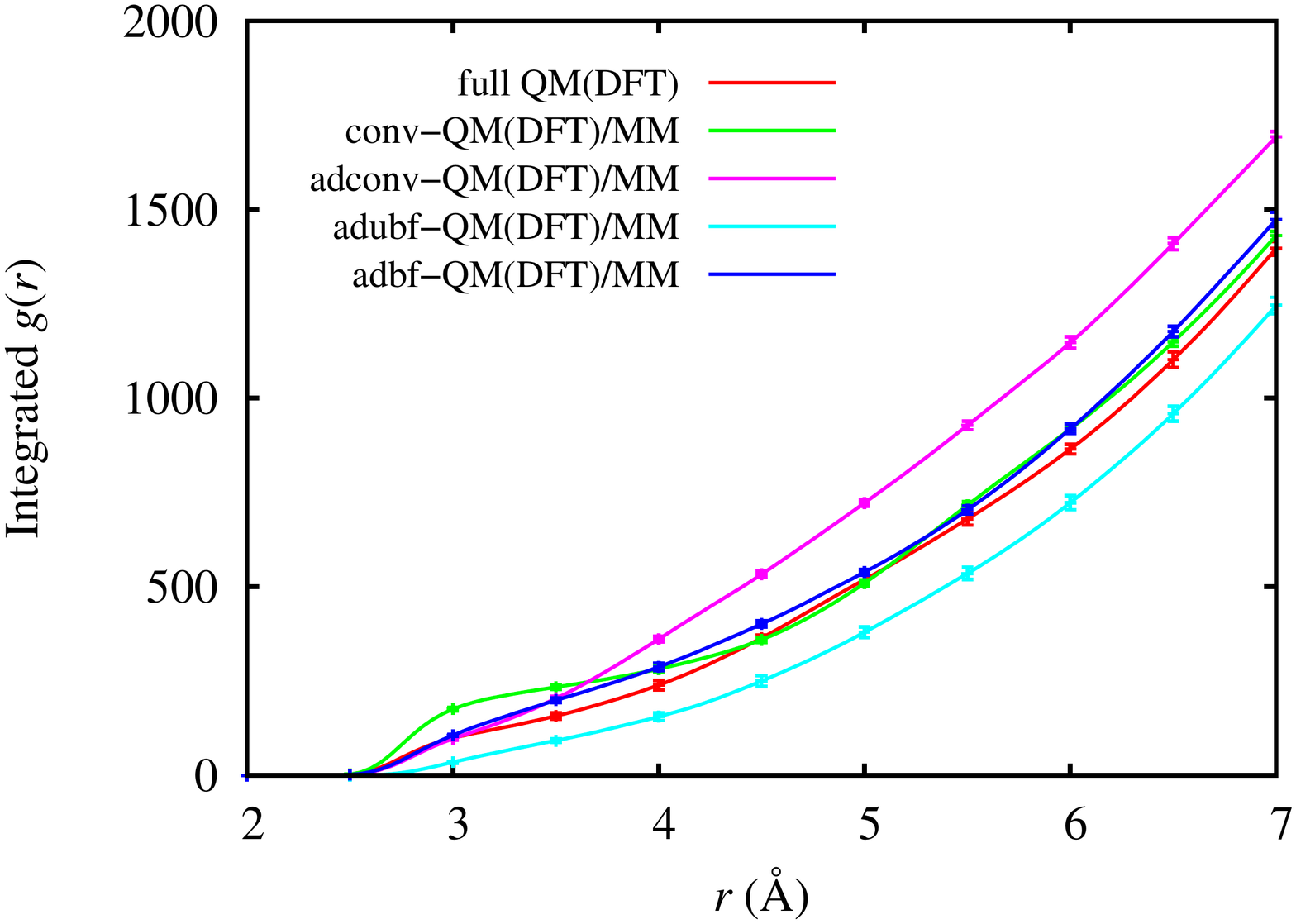}
        \caption{RDFs and integrated RDFs of bulk water using DFT with different adaptive QM/MM methods.}
        \label{fig:bulk_cp2k_rdf}
\end{figure}

In Fig.~\ref{fig:bulk_cp2k_rdf} we compare the O-O RDFs for DFT calculations
using fully QM, conv-QM/MM, adconv-QM/MM, adubf-QM/MM, and adbf-QM/MM.
All but adubf-QM/MM have a first neighbour peak at approximately the
correct distance, but their heights vary greatly.  In the conv-QM/MM
calculation, where only a single water molecule is in the QM region,
the first neighbour peak height is approximately double the fully QM
value, indicating that inaccurate forces at the QM-MM interface are
greatly distorting the structure around the QM water.  In the
adconv-QM/MM calculation, where the size of the dynamical QM region is increased
using hysteretic radii of 3.0-3.5~{\AA}, the peak height is
greatly improved, but there is an excess of molecules just inside
the QM-MM interface, leading to an unphysical second broad peak in
the RDF centered around about 3.8~{\AA}.  In contrast, using force
mixing without buffers in the adubf-QM/MM calculation leads to an
emptying of the dynamical  QM region, nearly completely eliminating the first
neighbour peak.  The adbf-QM/MM method comes closest to reproducing
the fully QM structure.  The first neighbour peak has the the correct
position and height, although the minimum near 3.2~{\AA} has
been replaced by a shoulder.  This artifact may be caused by the
nearby QM-MM interface, and could perhaps be corrected by a larger dynamical
 QM region.  Note that the effect is already much less significant
than the artifacts in the other adaptive methods.  The cumulative
RDFs in the bottom panel of Fig.~\ref{fig:bulk_cp2k_rdf} show corresponding
differences between the methods.  The conv-QM/MM curve shows a
large bulge near the first peak, but then follows the fully QM
curve at longer distances due to an overly deep minimum in the RDF that
compensates for the excess first neighbours.  The two unbuffered
adaptive methods show significant deviations from fully QM,
up for adconv-QM/MM which has an excess second RDF peak, and
down for adubf-QM/MM which is missing the first peak.  Our
adbf-QM/MM results show better agreement with fully QM throughout
the distance range, with a small offset to larger values starting
after the first RDF peak due to the shoulder in the peak.

\subsection*{\sffamily \large Water autoprotolysis in the presence of a zinc ion}

In the simulation of water autoprotolysis
in the presence of a Zn$^{2+}$ ion with the conv-QM/MM method, the QM region consists of the
metal ion and the reactant water molecules.  No additional water
molecules from the zinc's coordination sphere are included because
they are mobile, i.e. can exchange with bulk phase water on the simulation time scale, and the conv-QM/MM method is not adaptive.  A
possible way to keep these waters in the dynamical QM region is to restrain
them near the zinc ion\cite{Caratzoulas11}, or restrain the
remaining waters away from the dynamical  QM region~\cite{Rowley12}.
However, such restraints can significantly affect the entropic part of the
free energy\cite{Mones12} preventing the correct comparison of
the free energy profiles of the different methods.

\begin{figure}
        \centering
        \includegraphics[width=0.5\columnwidth]{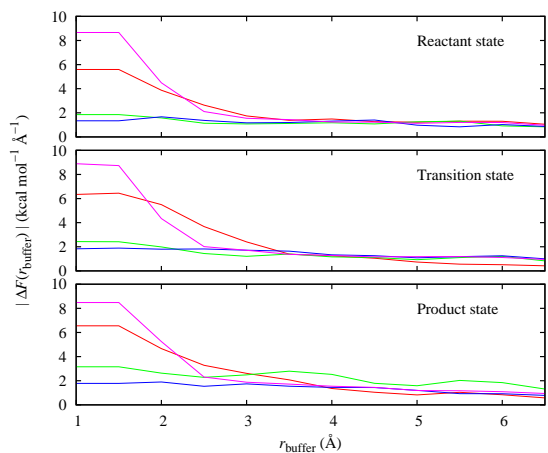}
        \caption{Mean force errors of key atoms in the dynamical QM region ($r_{\mathrm{qm}} = 3.0$~{\AA}) of the water autoprotolysis reaction using the MNDOd model and different sizes of buffer region at the three conv-QM/MM predicted extremum points, relative to forces from a calculation with buffer size of 7.0~{\AA}. Force errors on zinc ion (red), donor (green) and acceptor (blue) oxygen atoms and the average of non-reactive oxygen atoms (purple) in the dynamical QM region are shown.}
        \label{fig:auto_force_conv}
\end{figure}

For the adaptive QM/MM simulations we found that $r_\mathrm{qm} = 2.5-3.0$~{\AA} was sufficient to
include the first hydration shell around the zinc ion and the
reactants. To obtain the values of $r_\mathrm{buffer}$ we carried
out force convergence tests at geometries taken from the free energy profile extremum
states (reactant, transition and product) from the conv-QM/MM simulation.
The average and maximum force errors of the zinc, the donor and
acceptor oxygen atoms (which together comprise the core region) 
and the  oxygen atoms of non-reacting water molecules in the dynamical QM region 
are plotted in Fig.~\ref{fig:auto_force_conv}.  We see that including the
first hydration shell around the reactant water molecules is
sufficient to reduce the force error on all atoms to below approximately
2.5~kcal~mol$^{-1}$~{\AA}$^{-1}$, which we take to be an acceptable value.  Similarly
to bulk water, the hydrogen atoms have a slightly
faster convergence (data not shown). Interestingly, force errors
on the metal ion require $r_\mathrm{buffer} \ge 3.0$~{\AA} to
reach equally small values, despite the fact that it is
surrounded by QM waters in its first coordination sphere even without the use of a buffer region.  
The reason for this slow convergence is probably due to the metal ion's high charge and polarizability,
which cannot be fully screened by the coordinated water molecules. Based
on the convergence of the force on metal ion and the non-reactive water
molecules in the dynamical QM region we chose $r_\mathrm{buffer} = 3.0
- 3.5$~{\AA}.  Since the force convergence test showed small errors
on the reactants' atoms (although not on the metal ion, which functions
as a catalyst, not a reactant) even in the absence of any buffer
region, it may be reasonable to carry out the simulations without
a buffer.

We therefore also performed adconv-QM/MM
and adubf-QM/MM simulations using our AMBER implementation.

\begin{figure}
        \centering
                \includegraphics[width=0.5\columnwidth]{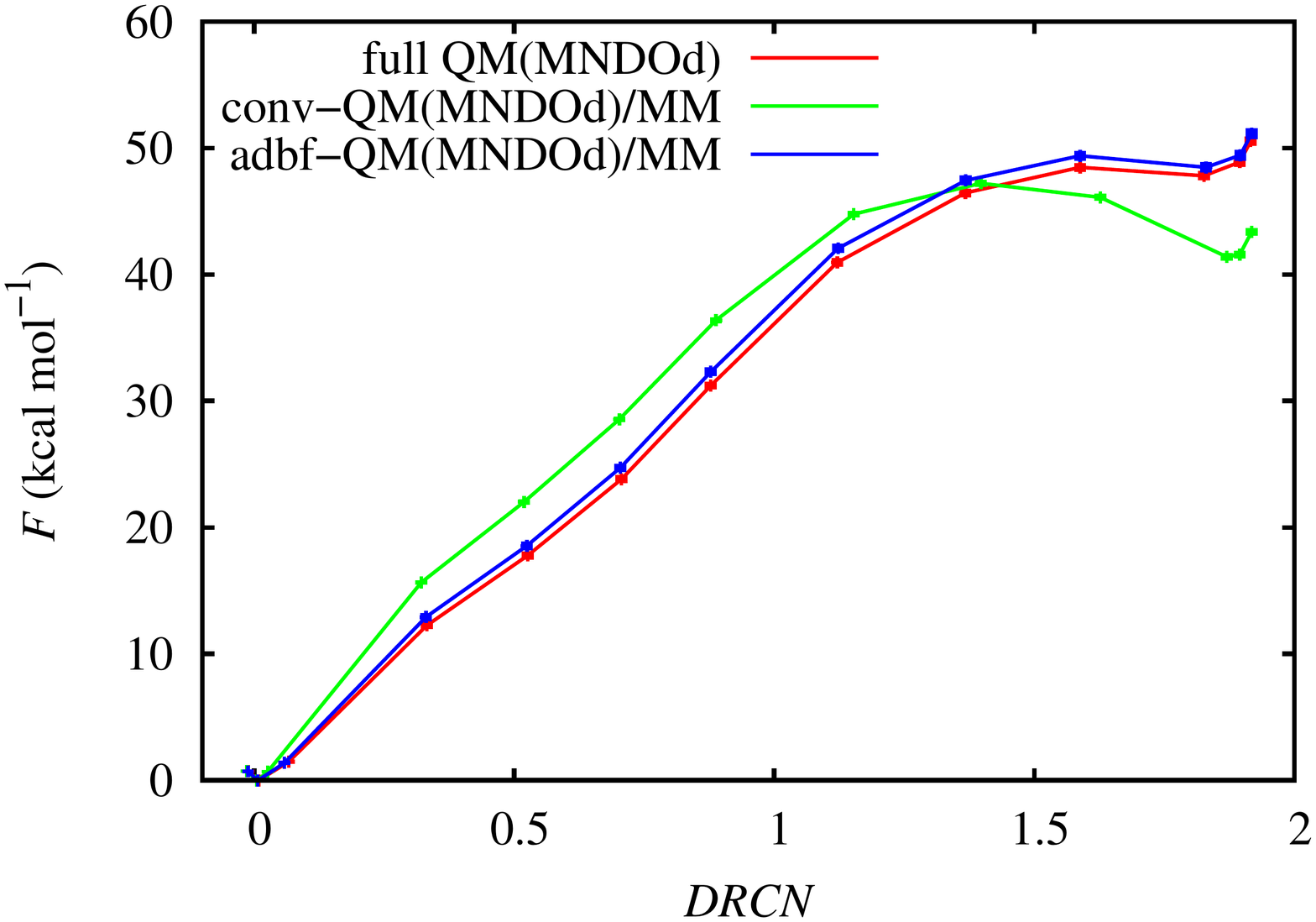}
                \includegraphics[width=0.5\columnwidth]{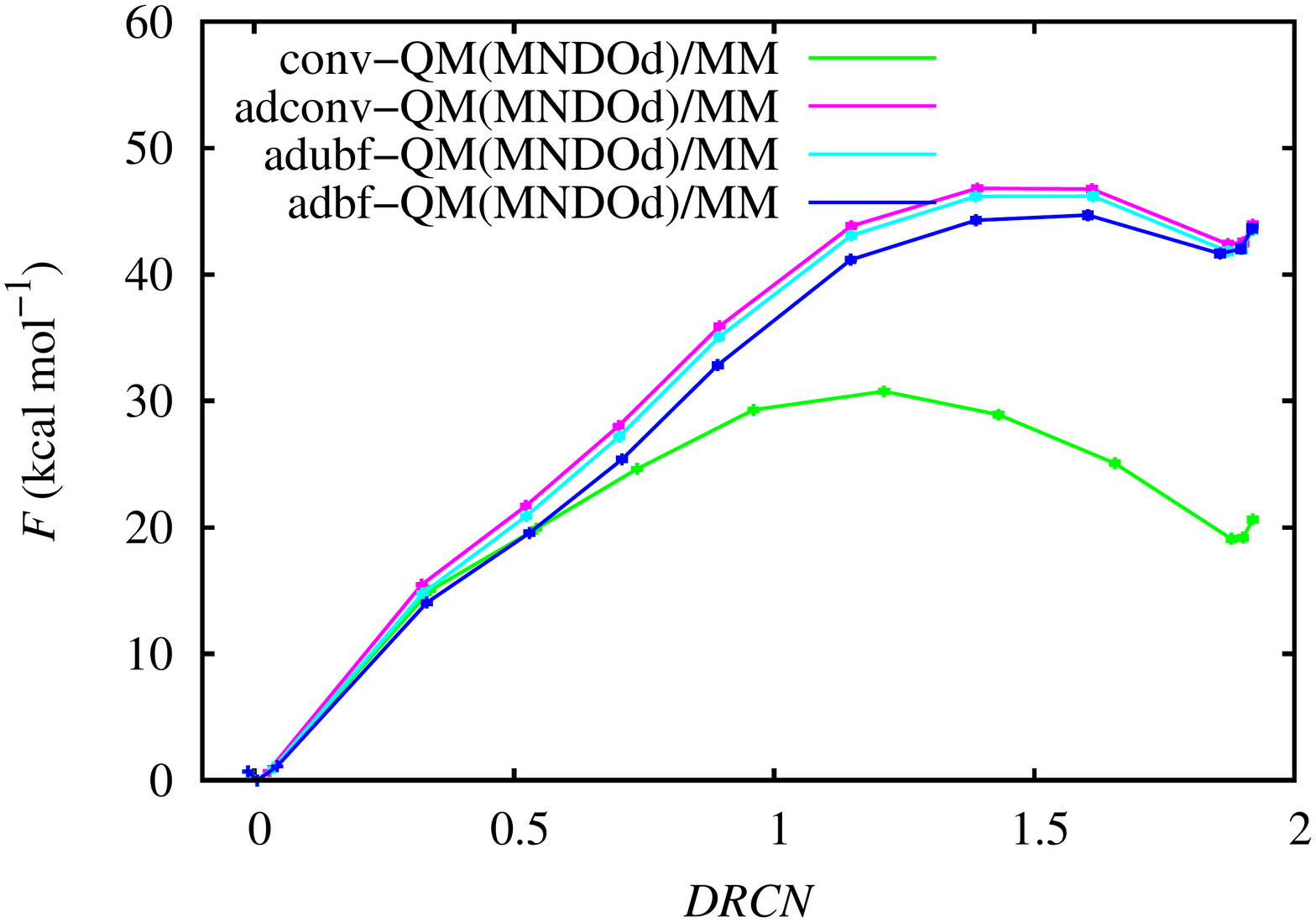}
        \caption{Potential of mean force profiles of the water autoprolysis reaction using MNDOd and the different QM/MM methods as functions of the difference of rational coordination number \emph{DRCN}. 95\% confidence intervals are comparable in size to symbols. Top panel shows results from CP2K including periodic fully SE simulation, and bottom panel shows results from AMBER.}
        \label{fig:auto_pmf}
\end{figure}

The free energy profiles of the different adaptive QM/MM methods calculated with the
CP2K and AMBER implementations are presented in Fig.~\ref{fig:auto_pmf}.
Since the formulations of the QM-MM interaction differ for the two
programs, we show the corresponding profiles in different figures.
For all cases \emph{DRCN} $ = 0.0$ corresponds to the reactant state. 
We used the profile of the smaller fully QM unit cell
size as reference, but as the larger fully QM unit cell size profile differs by
less than 0.025~kcal~mol$^{-1}$ RMS, we conclude that the small QM unit cell
profile is converged with respect to the unit cell size.
The curve of the fully QM simulation indicates the transition state (TS)
at around \emph{DRCN} $ = 1.6$ with an activation barrier of 48.5~kcal~mol$^{-1}$ 
and a shallow minimum of the product state at \emph{DRCN} $ \sim 1.8$ 
with a reaction free energy of 47.8~kcal~mol$^{-1}$.

As the reaction proceeds the conv-QM/MM profile diverges from the
rest. However, the deviation is much larger for the AMBER implementation
than for CP2K. This is probably due to  the differences in calculating the QM-MM interaction in the two programs; for example, the replacement of the point charges used in AMBER by Gaussians in CP2K may be reducing
the overpolarization of the QM calculation by the MM region and
leading to an improvement of the conv-QM/MM calculation.  In contrast
to conv-QM/MM, all of the adaptive methods accurately reproduce 
the fully QM results.  The adconv-QM/MM and adubf-QM/MM profiles differ
only slightly from the adbf-QM/MM one, in accord with the observation
of the force convergence test, where a QM region that included
the first hydration shell was sufficient to get force convergence
on the atoms of the reactants, even without the use of a buffer
region.

\subsection*{\sffamily \large Dimethyl-phosphate hydrolysis}

\begin{figure}
        \centering
        \includegraphics[width=0.5\columnwidth]{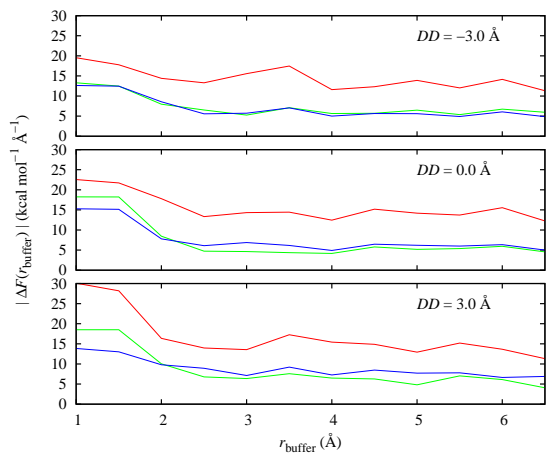}
        \caption{Mean force errors of key atoms of the phosphate hydrolysis reaction using DFT and different sizes of buffer region around the phosphate -- hydroxide ion system (i.e.\ $r_{\mathrm{qm}} = 0.0$~{\AA}) at three different \emph{DD} values, relative to reference forces from a calculation with buffer size of 7.0~{\AA}. Force errors on phosphorus atom (red), attacking (green) and leaving (blue) oxygen atoms are shown.}
        \label{fig:dmpoh_force_conv}
\end{figure}

To determine the sizes of the qm and buffer regions we carried out force convergence tests of 
the three key atoms of the system that are involved in the reaction coordinate \emph{DD}: the phosphorus atom and
the attacking and leaving oxygen atoms. First, we examined the effect of different buffer region sizes
directly around the phosphate and hydroxide ions by varying $r_{\mathrm{buffer}}$ with $r_{\mathrm{qm}} = 0.0$~{\AA} (Fig.~\ref{fig:dmpoh_force_conv}).
In general we found a much slower force convergence as compared to the water autoprotolysis, which can be explained by the highly negatively charged
species in this system.
For the oxygen atoms a similar behaviour of the profiles can be seen
for all three \emph{DD} values we investigated: without a buffer (i.e.\ $r_{\mathrm{buffer}} = 1.0$~{\AA}, which is too
small to include any neighboring molecules) the error 
is about 15-20~kcal~mol$^{-1}$~{\AA}$^{-1}$, and it goes down to 5-6~kcal~mol$^{-1}$~{\AA}$^{-1}$ using a buffer size of 3.0-3.5~{\AA}. 
This buffer size corresponds to the first hydration shell around the reactants, and
applying a larger buffer size does not improve the force convergence. 
For the phosphorus atom the force convergence profile shows a similar behaviour but converges 
to a larger average force error of $\sim 15$~kcal~mol$^{-1}$~{\AA}$^{-1}$.
Based on Fig.~\ref{fig:dmpoh_force_conv} we set $r_{\mathrm{qm}} = 3.0-3.5$~{\AA}.
We also investigated the convergence of forces as a function of buffer region 
around a finite dynamical QM region ($r_{\mathrm{qm}} = 3.0-3.5$~{\AA}). In this case we did not find any additional improvement
of the force convergence, which is in agreement with the tail of the profiles in Fig.~\ref{fig:dmpoh_force_conv} and suggests that applying a buffer
region beyond the dynamical  QM region that includes the first hydration shell will not alter the free energy profile significantly.
We tested this by using $r_{\mathrm{buffer}} = 3.0-3.5$~{\AA}, as in the other simulated systems, in the adbf-QM/MM calculations.

\begin{figure}
        \centering
        \includegraphics[width=0.5\columnwidth]{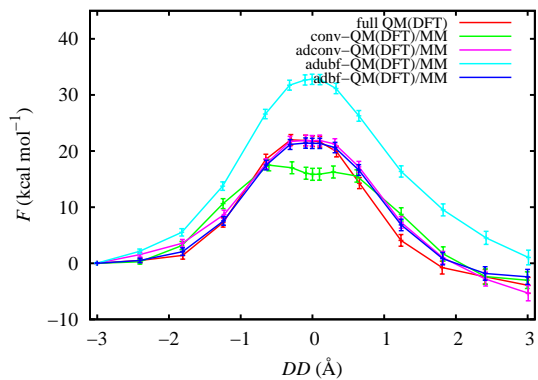}
        \caption{Potential of mean force profiles of the phosphate hydrolysis reaction using DFT and the different adaptive QM/MM methods as functions of the distance 
        difference \emph{DD}, with 95\% confidence intervals.}
        \label{fig:dmpoh_pmf}
\end{figure}

The free energy curves of the conventional and adaptive QM/MM simulations of the
systems are shown in Fig.~\ref{fig:dmpoh_pmf}. All profiles are shifted to $F=0~$kcal~mol$^{-1}$ at \emph{DD}$= -3.0$~{\AA}. The fully QM profile has a maximum at \emph{DD}$= -0.3$~{\AA}
with $\Delta F^{\ddagger} = 22.0$~kcal~mol$^{-1}$,
indicating the transition state of the
reaction. Within the range of the UI calculations ($[-3.0, 3.0]$~{\AA})
the fully QM profile does not have minima as expected due to the
repulsion of the negatively charged reactants and products.  The
conv-QM/MM simulations result in a wide flat region in the range
$[-0.6, 0.6]$~{\AA} with a minimum at \emph{DD}$= 0.0$~{\AA}, indicating
a possible intermediate metastable state rather than a transition state, 
although the observed minimum is shallower than the error bars.
The top of the conv-QM/MM profile is lower
by $\sim 5$~kcal~mol$^{-1}$ than the peak of the fully QM curve, 
corresponding to an error of $\sim 25\%$.  The adubf-QM/MM profile
has a single well defined transition state, but its height is
significantly overestimated compared to fully QM ($\Delta F^{\ddagger} =
32.8$~kcal~mol$^{-1}$).  In contrast to conv-QM/MM and adubf-QM/MM, both
the adconv-QM/MM and adbf-QM/MM profiles are in good agreement with the
fully QM profile.  

\begin{figure}
        \centering
        \includegraphics[width=0.5\columnwidth]{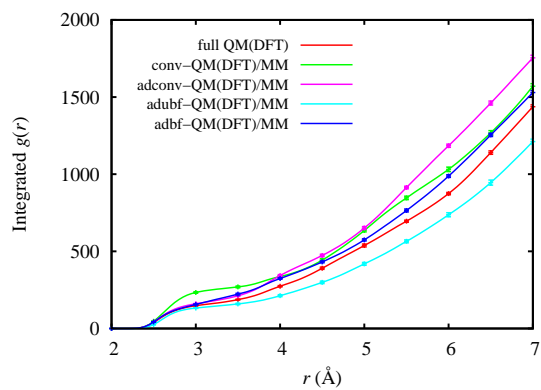}
        \caption{Integrated central phosphorus -- water oxygen RDF at the transition state corresponding to the fully QM simulation of the phosphate hydrolysis reactions using different adaptive QM/MM methods, with 95\% confidence intervals.}
        \label{fig:dmpoh_rdf}
\end{figure}

To further investigate the source of these differences, we computed
the RDFs between the central phosphorous and all water oxygen atoms.
Instead of using the RDF, which is noisy
due to the relatively short simulation times, we calculated its
integral (IRDF), shown in Fig.~\ref{fig:dmpoh_rdf}.  For smaller
distances (2.5-3.0~{\AA}) the conv-QM/MM IRDF shows a higher water
density around the reactants compared to the fully QM simulations.
This is due to the ability of the MM hydrogen atoms to approach the
pentavalent transition state too closely, leading to an overstabilization of the 
doubly negatively charged phosphate and resulting in a lower barrier.  In
the case of adubf-QM/MM the IRDF profile shows that an instability has
pushed water molecules out of the dynamical QM region, decreasing
the density for $r$ at least up to 7~{\AA}.  This unphysically low
density in the reaction region reduces the stabilization of the
transition state by the nearby waters, in accord with the
higher barrier observed.  In the vicinity of the reactants both the
adconv-QM/MM and adbf-QM/MM integrated RDFs are close to the fully QM
one.  At larger distances (starting from 4~{\AA}) the adconv-QM/MM
RDF starts to diverge while adbf-QM/MM remains closer to the fully QM
result, although the adconv-QM/MM method's structural error does
not significantly affect its free energy profile.

\section*{\sffamily \Large DISCUSSION\ AND\ CONCLUSIONS}

The QM/MM approach has been widely used for simulating processes
that require a quantum-mechanical description in a small region,
for example a reaction with covalent bond rearrangement, within a
larger system with important long-range structure, such as a protein
or a polar solvent.  However, conventional approaches are limited to a
fixed QM region, and also contain significant errors in atomic forces near the QM-MM
interface as compared with fully QM or fully MM simulations.  Making
the QM region larger can help by moving the QM-MM interface further away
from the region of interest, but may require the methods to
become adaptive by allowing molecules to diffuse into or out of
the QM region.  Such adaptive methods have
been developed, but it has proven difficult to make them stable,
at least partially because the force errors near the QM-MM interface
can unphysically drive particles from one region to the other.  To
address these issues and enable stable adaptive simulations we have developed the adbf-QM/MM method, which
reduces interface errors by combining forces from two QM/MM calculations with different QM sizes  using force-mixing.  Here we have described
its implementation in the CP2K and AMBER programs, building on their
existing  QM/MM capabilities.  Using the new functionality
requires the specification of  a few parameters to
control the sizes of the core QM, dynamical QM, and buffer regions.  The adbf-QM/MM
method and its implementations are formulated in a general way, so
they can be used with a wide range of QM and MM models as well as
different  QM/MM coupling methods.

We have tested our implementations using a variety of QM models,
including both semi-empirical and density functional theory, on several structural and free energy
problems, using conventional QM/MM, adbf-QM/MM, as well as other
adaptive methods that forgo the use of some of the QM and/or MM buffer
regions.  Using the CP2K and AMBER implementations we simulated the structure
of bulk water, where we have shown that adbf-QM/MM produces a stable
structure in good agreement with fully QM simulations for DFT and for some, but
not all, SE methods we tested.  A comparison of the free energy
profiles of two reactions, water autoprotolysis in the presence of a
Zn$^{2+}$ ion (SE using AMBER) and dimethyl-phosphate hydrolysis (DFT using CP2K), to fully QM 
results shows a substantial dependence on the choice of
adaptivity, buffers, and details of the QM-MM interaction term.  In
all cases, the use of a simulation that includes at least one
hydration shell beyond the reacting species is important for
reproducing the fully QM free energy profile.  The water autoprotolysis
simulations show some differences between AMBER and CP2K due to their
differing QM-MM interactions, but the adbf-QM/MM method gives good
agreement with fully QM simulations for both software packages despite these differences.  The dimethyl phosphate
hydrolysis simulations show that the free energy profiles of the 
adconv-QM/MM and adbf-QM/MM adaptive method are in good agreement with fully
QM results, while the conv-QM/MM and adubf-QM/MM methods are not.  
The reason for this difference is that the former two methods result
in a reasonable solvent structure around the reaction, while the latter
two give very different structures.  The conv-QM/MM simulation also
 predicts a qualitatively incorrect metastable state at the
transition state collective coordinate value.  

In summary, our results show that of the adaptive methods we have tested, the adbf-QM/MM method is the most robust 
in maintaining reasonable solvent structure
and giving accurate free energy profiles.  
Adaptive methods that do not include both dynamical QM
and buffer regions can also give good structural and free energy profile results 
for some systems, but they fail to agree with full QM results for other systems.
To maximize the accuracy of the adbf-QM/MM method the size of core
 region should be minimized, the dynamical QM region should
include at least one hydration shell around the reaction centre so as to include the
most important solvent effects, and the buffer region should be large
enough to give  forces throughout the dynamical QM region converged to better than a few kcal mol$^{-1}$ \AA$^{-1}$.
Our AMBER and CP2K implementations  use a small number
of simple parameters to specify the various adaptive regions, and
the suggested size criteria can be satisfied with reasonable computational
cost, making the adbf-QM/MM method accessible to a wide community of
users.

\subsection*{\sffamily \large ACKNOWLEDGMENTS}

N.B. acknowledges funding for this work by the Office of Naval Research through the Naval Research Laboratory's basic research program, and computer time at the AFRL DoD Supercomputing Resource Center through the DoD High Performance Computing Modernization Program (subproject NRLDC04253428). B.L. was supported by EPSRC (grant no. EP/G036136/1) and the Scottish Funding Council. G.C. and B.L. acknowledge support form EPSRC under grant no. EP/J01298X/1. R.C.W. and A.W.G. acknowledge financial support by the National Institutes of Health (R01 GM100934), A.W.G. acknowledges financial support by the Department of Energy (DE-AC36-99GO-10337). This work was partially supported by National Science Foundation (grant no. OCI-1148358) and used the Extreme Science and Engineering Discovery Environment (XSEDE), which is supported by National Science Foundation grant no. ACI-1053575. Computer time was provided by the San Diego Supercomputer Center through XSEDE award TG-CHE130010. 

\clearpage

\bibliography{abfqmmm_paper}

\clearpage

\listoffigures

\newpage

\clearpage

\end{document}